\documentclass[10pt,twocolumn,letterpaper]{article}

\usepackage{amsfonts}
\usepackage{amsmath}
\DeclareMathOperator{\Tr}{Tr}
\usepackage{mathabx}					 	
\usepackage{placeins}                       
\usepackage{subcaption,caption}
\usepackage{epsfig}
\usepackage{booktabs}                       
\usepackage{tabularx}                       
\graphicspath{{Figures/}}                   
\usepackage{accents}                        
\usepackage{supertabular}
\usepackage{multirow}
\usepackage{longtable}
\usepackage{capt-of}                        
\usepackage{comment}                        
\usepackage{algorithm}                   
\usepackage{algpseudocode}
\usepackage[hyphens]{url}
\usepackage[colorlinks=true,urlcolor=blue]{hyperref} 

\usepackage{iccv}
\usepackage{times}
\usepackage{graphicx}
\usepackage{amssymb}
\usepackage{bbm}
\usepackage{mathrsfs}

\iccvfinalcopy 

\def\iccvPaperID{406} 

\begin{document}

\title{Calibration of the von Wolffersdorff model using Genetic Algorithms}

\author{Francisco J. Mendez{$^{1,*}$},~ Antonio Pasculli{$^{1}$},~Miguel A. Mendez{$^{2}$},~ Nicola Sciarra{$^{1}$}\\\vspace{-8pt}{\small~}\\
{$^{1}$}{\small University G. D'Annunzio, Dept. of Engineering Geology (INGEO), Chieti-Pescara, Italy};\\ {$^{2}$} {\small von Karman Institute for Fluid Dynamics, EA Department, Sint-Genesius-Rode, Belgium}\\
{\small{{$^{*}$}Corresponding to: \tt{francisco.mendez@unich.it}}}}


\maketitle
\thispagestyle{empty}

\begin{abstract}
This article proposes an optimization framework, based on  Genetic Algorithms (GA), to calibrate the constitutive law of von Wolffersdorff. This constitutive law is known as Sand Hypoplasticity (SH), and allows for robust and accurate modeling of the soil behavior but requires a complex calibration involving eight parameters. The proposed optimization can automatically fit these parameters from the results of an oedometric and a triaxial drained compression test, by combining the GA with a numerical solver that integrates the SH in the test conditions.  
By repeating the same calibration several times, the stochastic nature of the optimizer enables the uncertainty quantification of the calibration parameters and allows studying their relative importance on the model prediction. After validating the numerical solver on the ExCaliber-Laboratory software from the SoilModels' website, the GA calibration is tested on a synthetic dataset to analyze the convergence and the statistics of the results. In particular, a correlation analysis reveals that two couples of the eight model parameters are strongly correlated. Finally, the calibration procedure is tested on the results from von Wolffersdorff, 1996, and Herle \& Gudehus, 1999, on the Hochstetten sand. The model parameters identified by the Genetic Algorithm optimization improves the matching with the experimental data and hence lead to a better calibration.
\end{abstract} 


\section{Introduction}
To model the mechanical behavior of the soil, a large variety of constitutive laws have been developed, among which the hypoplasticity\cite{Lade_Overview_of_Constitutive_Models_For_Soils,Masin_Modelling_of_Soil_Behaviour_with_Hypoplasticity} .This term has been coined in the 1986 by Dafalias \cite{Dafalias_Bounding_Surface_Plasticity_Mathematical_Foundation_and_Hypoplasticity} although the first constitutive law has been proposed in the 1977 by Kolymbas \cite{Kolymbas_A_rate_dependent_constitutive_equation_for_soils}. The study of the hypoplastic equations has been pioneered in the University of Karlsruhe and Grenoble, with the objective of developing constitutive models for granular materials, such as sand and gravel \cite{Tamagnini_Viggiani_Chambon_A_review_of_two_different_approaches_to_hypoplasticity}. Even if its earliest formulation did not take into account the void ratio as a state variable, the hypoplastic equations proved to be a powerful tool to describe the mechanical behavior of the soil \cite{Wei_Kolymbas_Numerical_testing_of_the_stability_criterion_for_hypoplastic_constitutive_equations}. 

Later, combining the contributions of Gudehus \cite{Gudehus_A_comprehensive_}
and Bauer \cite{Bauer_Calibration_of_a_comprehensive_hypoplastic_model_for_granular_materials}, 
von Wolffersdorff formulated the set of equations that summarized over 25 years of previous studies \cite{Wolffersdorff_A_hypoplastic_for_granular_material_with_a_predefined_limit_state_surface}.
In this latest version, often referred to as Sand Hypoplasticity (SH), the constitutive law was able to describe the soil dilatancy, pyknotropy, barotropicity, and the critical state. This model, although not without issue \cite{Wu_A_basic_hypoplastic_constitutive_model_for_sand}, has stood the test of time and remains an important tool to describe soils composed of undeformable and cohesionless grains.

For cohesive soils, on the other hand, different kinds of hypoplastic constitutive laws have been developed. Nemunis proposed a visco-hyploplastic approach \cite{Wu_Bauer_Niemunis_Herle_A_visco-hypoplastic_model_for_cohesive_soils,Niemunis_Grandas-Tavera_Prada_Anisotropic_visco-hypoplasticity}, while Ma\v{s}\'in elaborated an hypoplastic version of the Cam Clay model \cite{Masin_Clay_hypoplasticity_model_including_stiffness_anisotropy,Masin_A_hypoplastic_constitutive_model_for_clays,Masin_Clay_hypoplasticity_with_explicitly_defined_asymptotic_states}. With the intergranular strain concept proposed by Niemunis \& Herle \cite{Niemunis_Herle__Hypoplastic_model_for_cohesionless_soils_with_elastic_strain_range}, which allows for extending the SH model to small deformation and cyclic loads, the hypoplastic theory is today able to describe the behaviour of granular soil in a wide range of geotecnical problems \cite{info_sol_model_projecr,A_hypoplastic_model_for_site_response_analysis_KReyes,Effects_of_Pillar_Depth_and_Shielding_on_the_Interaction_of_Crossing_Multitunnels_Ng_C_W_W}. The interest of the scientific and professionals community for this family of constitutive laws is proven by their large diffusion in commercial codes, among which \textsc{Abacus}, \textsc{Diana} and \textsc{Plaxis}. 

This article focuses on one of the most delicate aspects in the use of the SH constitutive law: its calibration, which is the identification of the model parameter for a given soil. This model depends on eight interconnected parameters that govern a strongly nonlinear dynamic system.  One of the most important contributions to the calibration of the SH model was proposed in 1996 by Herle \& Gudehus  \cite{Herle_Gudehus_Determination_of_parameters_of_a_hypoplastic_constitutive_model_from_properties_of_grain_assemblies}. These authors have derived the analytic equations for estimating the SH parameters, and defined the experimental procedures required for their identification. However, some of these analytical formulae are extremely sensitive to the input data and can lead to considerable uncertainties in the estimated parameters. Moreover, the calibration procedure proposed by Herle \& Gudehus requires laboratory analyses that are uncommon in practice, which is usually limited to the triaxial and eodometric compression tests.

To calibrate the SH model relying only on the results of these two tests, an optimization procedure is required. A free tool that for such calibration has been developed by T. Kadl\'i\v{c}ek, T. Janda and M. \v{S}ejnoha \cite{Kadlicek_Calibration_of_Hypoplastic_Models_for_Soils,Kadlicek_Automatic_online_calibration_software_excalibre} and is available at \url{soilmodels.com/excalibre/}. However,  this procedure tends to suppress the dilatancy and requires manual adjustments of some of the parameters.

The scope of this work is to present an approach that returns all the SH parameters with no need for manual adjustments. This approach is based on a Genetic Algorithm (GA) optimizer, which interacts with a fast solver for the SH model to reproduce the results of the eodomeric and triaxial compression tests. 

A fundamental tool on which GA is based is the generation of pseudorandom numbers. This type of approach is commonly used by the  Monte Carlo method which has been also usefully applied in different fields, for example: \cite{Calista_PASCULLI,PASCULLI2018370}. The GA has been initially developed by Holland \cite{Holland1992} in 1975 and later popularized by the excellent book of Goldberg \cite{1989a}. This algorithm is a global minimum optimizer, inspired by the principles of genetics and natural selection. In the era of the big data revolution, the GA has become a fundamental tool in a wide range of applications, including operation management \cite{Lee2018}, image reconstruction \cite{Mirjalili2019}, data-driven control \cite{Duriez2017} and Machine Learning \cite{Shapiro2001}. The significant advantage of GA is easy programming and parallelization. Moreover, the GA offers a good balance between fast convergence and exploratory search, allowing for escaping from local minima and aiming to the global one. An excellent introduction to the subject is the monographs from Haupt \& Haupt \cite{Haupt2003} and Michalewicz \cite{Michalewicz1996}.

In this work, the GA optimizer operates on the set of model parameters, comparing the corresponding numerical prediction of the SH model to the experimental results until the best set is identified. The model equations implemented in the SH solver are described in Section \ref{Par2}, including both the general formulation and the simplified forms involved in the specific tests considered in this work. Section \ref{Tre} describes the calibration methodology, including the integration procedure, the treatment of the different tests, the formulation of the cost function to minimize, and the GA optimizer. The results are presented in Section  \ref{Quattro}, which is divided into three parts. Section \ref{subpar:Validazione} presents a validation of the integration procedure. Section \ref{subpar:ConvergenzaMetodo} focuses on the problem of solution uniqueness and its link to the sensitivity of the model and the uncertainty of the identified coefficients. These points are addressed by using the numerical model to construct synthetic experimental data and then testing the capabilities of the optimizer to retrieve the coefficients from which the data is generated.
Finally, in Section \ref{subpar:Calibrazione}, the algorithm is tested on the experimental dataset provided by von Wolffersdorff in \cite{Wolffersdorff_A_hypoplastic_for_granular_material_with_a_predefined_limit_state_surface}. The calibration from GA, von Wolffersdorff and Herle \& Gudehus 
\cite{Herle_Gudehus_Determination_of_parameters_of_a_hypoplastic_constitutive_model_from_properties_of_grain_assemblies} are compared. The conclusions are collected in Section \ref{Conc}.

\section{The Sand Hypoplasticity (SH) Model}\label{Par2}

The Sand Hypoplasticity (SH) theory considers the soil as a continuous porous media for which it is possible to define a constitutive law in terms of rate-equations \cite{Kolymbas_Introduction_to_Hypoplasticity}. These rate-equations represent a nonlinear dynamical system describing the time evolution of the objective stress tensor $\accentset{\circ}{\mathbf{T}}\in \mathbb{R}^{3\times3}$ to the granulate stretching rate $\mathbf{D}= (\nabla v_s + \nabla v_s^T)/2\in \mathbb{R}^{3\times3}$, where $v_s$ is the velocity of the grain skeleton, the Cauchy effective stress ${\mathbf{T}}$, and the void ratio $e$:

\begin{equation}
\label{Eq1}
\begin{cases} 
\accentset{\circ}{\mathbf{T}}=\mathbf{F}\bigl(\mathbf{D},\mathbf{T},e,\mathbf{P}\bigr)\\ 
\dot{e}= g (\mathbf{D},e)
\end{cases}
\end{equation}

The objective stress tensor, used  to preserve the independence on the frame of reference, is defined following Zaremba-Jaumann \cite{Zaremba} as:  

\begin{equation}\label{eq:oT}
\accentset{\circ}{\mathbf{T}}=\dot{\mathbf{T}}-\mathbf{W}\cdot \mathbf{T}+{\mathbf{T}}\cdot{\mathbf{W}}\,,
\end{equation} where $\dot{\mathbf{T}}$ is the time derivative of the Cauchy effective stress and $\mathbf{W}=(\nabla v_s -\nabla v_s^T )/2$ is the spin tensor.

The nonlinear function $\mathbf{F}$ depends the set of eight parameters $\mathbf{P}\in \mathbb{R}^{8\times 1}$. The calibration procedure consists in identifying these parameters so that the solution of the dynamical system in  \eqref{Eq1} recovers the experimental results from two classical tests: the oedometer and the triaxial drained test. The function $g$ express the mass conservation of the sample during the test. Neglecting the deformability of the grains, this function relates the time evolution of the void ratio to the volumetric deformation as follows

\begin{equation}
\dot{e}=(1+e)\Tr\Bigl(\mathbf{D}\Bigr)\,.
\label{eq:e_dot}
\end{equation}

This section describes how to obtain the dynamical system in \eqref{Eq1}; section \ref{Tre} describes the optimization procedure to identify the model parameters. 

\subsection{General Formulation}

Following the formulation from von Wolffersdorff \cite{Wolffersdorff_A_hypoplastic_for_granular_material_with_a_predefined_limit_state_surface}, the nonlinear function $\mathbf{F}$ in \eqref{Eq1} becomes:

\begin{equation}
\accentset{\circ}{\mathbf{T}} = \frac{f_e\,f_b}{\Tr(\hat{\mathbf{T}}^2)}  \Bigl( F^2 \mathbf{D} + a^2\Tr(\hat{\mathbf{T}}\cdot \mathbf{D})\hat{\mathbf{T}}  +  f_d\,a\,F \bigr(\hat{\mathbf{T}}+\hat{\mathbf{T}}^*\bigl)\Vert \mathbf{D}  \Vert \Bigr)
\label{eq:Wolffersdorff}
\end{equation} where $\Tr$ denotes the trace of a tensor, $\Vert \, \Vert $ is the tensor norm $\Vert A \Vert=\sqrt{\Tr{(AA^T)}}$, $\hat{\mathbf{T}}^*= \hat{\mathbf{T}}-1/3 \mathbf{I}$, with  $\mathbf{I}$ the  identity tensor and $\hat{\mathbf{T}}={\mathbf{T}}/{\Tr({\mathbf{T}}})$. 

The coefficients $(f_e,f_b,F,a,f_d)$ have a semi-empirical interpretation and depend on the parameters of the model that needs to be tuned during the calibration. The coefficients $a$ and $F$ are linked to the critical yielding surface from Matsuoka-Nakai \cite{Matsuoka} and are computed as: 

\begin{align} \label{eq:a}
&a				=\frac{\sqrt{3}(3-\sin \varphi_c)}{2\sqrt{2}\sin \varphi_c }\\
&F				=\sqrt{\frac{1}{8}\tan^2\psi+\frac{2-\tan^2\psi}{2+\sqrt{2}\tan\psi \cos 3\theta}}-\frac{\tan \psi}{2\sqrt{2}} \,,
\end{align}\,where: 
\begin{align}
&\tan \psi		=\sqrt{3}\Vert \hat{\mathbf{T}}^*\Vert \\
&\cos 3\theta	=-\sqrt{6}\,\frac{tr(\hat{\mathbf{T}}^{*3} )}{\bigr[ \Tr(\hat{\mathbf{T}}^{*2})]^{3/2}} \label{5a}\,\,.
\end{align}
In the hydrostatic conditions (i.e., $\tan\psi =0$) and in axysiymmetric conditions (i.e.,
$0\leq \tan \psi\leq \sqrt{2}$ and $\cos 3\theta=-1$), 
the equation \eqref{5a} is an undetermined function tending to $F=1$ \cite{Wolffersdorff_A_hypoplastic_for_granular_material_with_a_predefined_limit_state_surface}.

The barotropy ad piknotropy coefficients $f_b$ and $f_e$ in \eqref{eq:Wolffersdorff} were originally formulated as:

\begin{align}
&f_b=\cfrac{\Biggr(\frac{e_{i0}}{e_{c0}}\Biggl)^\beta \,\cfrac{h_s}{n}\frac{1+e_i}{e_i}\Bigr(-\cfrac{\Tr(\mathbf{T})}{h_s}\Bigl)^{1-n}}{3+a^2-a\,\sqrt{3}\, \Biggr( \cfrac{e_{i0}-e_{d0}}{e_{c0}-e_{d0}}\Biggl)^\alpha}\,, \\
&f_e				=\Biggr(\frac{e_c}{e}\Biggl)^\beta\,. 
\end{align}

However, $f_e$ and $f_b$  are usually replaced by their product  $f_s$: 
 
 \begin{equation}
f_s=\cfrac{\cfrac{h_s}{n}\cfrac{1+e_i}{e_i}\Biggr(\cfrac{e_{i}}{e}\Biggl)^\beta \,\Biggr(-\cfrac{\Tr(\mathbf{T})}{h_s}\Biggl)^{1-n}}{3+a^2-a\,\sqrt{3}\, \Biggr( \cfrac{e_{i0}-e_{d0}}{e_{c0}-e_{d0}}\Biggl)^\alpha}
\label{eq:f_s}
\end{equation}

The coefficient $f_d$ is the pyknotropy coefficient defined as 

\begin{equation}
f_d=\Biggr(\frac{e-e_d}{e_c-e_d}\Biggl)^\alpha\,.
\label{eq:f_d}
\end{equation}

The previous equations depends on the maximum ($e_d$), minimal ($e_i$) and critical ($e_c$) void fractions. These are linked, according to Bauer 
\cite{Bauer_Calibration_of_a_comprehensive_hypoplastic_model_for_granular_materials}, by the the system:
 
\begin{equation}\label{eq:ei_ed_c_ed}
\frac{e_i}{e_{i0}}=\frac{e_d}{e_{d0}}=\frac{e_c}{e_{c0}}=\exp\Biggr[-\Biggr(\frac{-\Tr(\mathbf{T})}{h_s}\Biggl)^n\Biggl]\,.
\end{equation}

\begin{figure}[h]
\centering
\includegraphics[width=0.45\textwidth]{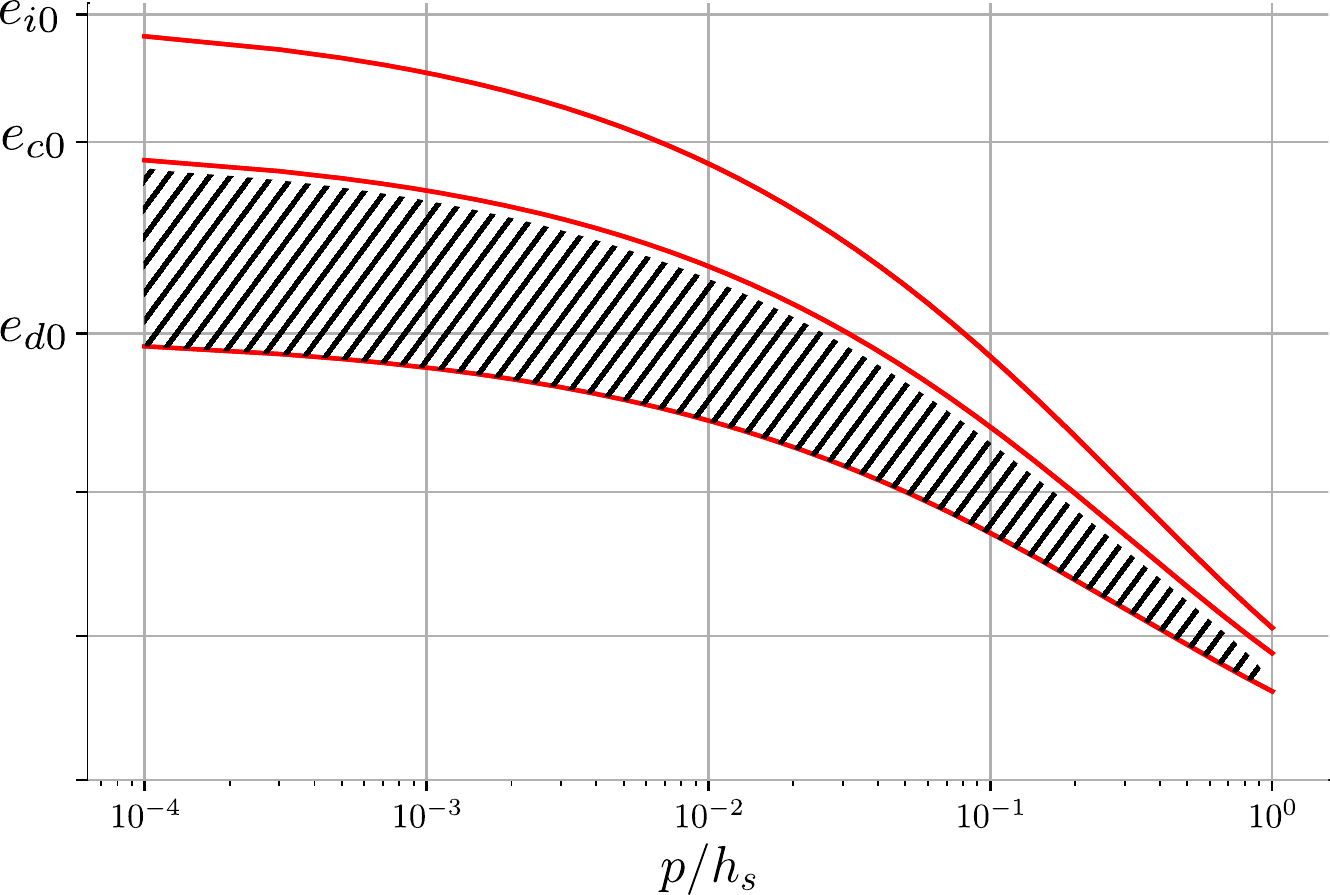}
\caption{Representation of Baure's laws \eqref{eq:ei_ed_c_ed} for $n = 0.4$ in the $p/h_s\,-\,e$ plane. The dashed area represents the region of dilative behaviour of the soil and the dotted area the contractive one. Re-adapted from \cite{Niemunis_Extended_hypoplastic_models_for_soils}.}
\label{fig:zone_di_comportamento}
\end{figure}

For a given mean pressure $p=-tr(\mathbf{T})/3$, among the possible void ratio $e_d<e<e_i$, we can identify regions of dilative (for $e_d<e<e_c$) and contractive (for $e_c<e<e_i$) behavior.
Figure \ref{fig:zone_di_comportamento} shows these two regions in the plane ($p/h_s\,, \,e$). 


Finally, the set of equations \eqref{eq:a}-\eqref{eq:ei_ed_c_ed} include eight unknown parameters $\mathbf{P}=\{e_{c0} ,  e_{d0} ,  e_{i0} ,  h_s ,  \phi ,  n ,  \alpha ,  \beta \}$, which are herein described \cite{Masin_The_influence_of_experimental_and_sampling_uncertainties_on_the_probability_of_unsatisfactory_performance_in_geotechnical_applications}.

\begin{itemize}
\item  $e_{c0}$,  $e_{d0}$  $e_{i0}$. These are, respectively, the critical the minimal and the maximal void ratios, obtained when $p_s=\Tr(\mathbf{T})=0 $ \eqref{eq:ei_ed_c_ed}. 
The ratios $\lambda_d=e_{d0}/e_{c0}$ and $\lambda_i=e_{i0}/e_{c0}$
govern the amplitude of the domains of dilatant or contractive behaviour, while $e_{c0}$ defines the critical state in terms of void ratio.

 \item $h_s$ is called granular hardness. This has the dimensions of $kPa$ but should not be confused with the grains rigidity which are considered undeformable. This parameter is linked to the barotropy of the solid skeleton and its increase of the slope of eodometric curve response.

\item $\varphi_c$ is the well-known critical friction angle and is linked to the  shear strength in critical conditions.

\item $n$ is a parameter influencing the barotropy of the soil. Increasing $n$ produces an increase of the curvature in the response curve in the eodometric tests. 

\item $\alpha$ is the exponent in the calculation of the picnotropy coefficient $f_d$ and controls the dependency of peak friction angle on relative density. 
 
\item $\beta$ is an coefficient influencing barotropy and pikno-tropy. Increasing $\beta$ produce an increase of the stiffness of material and in particular the shear stiffness.

\end{itemize}

The range of these parameters for various granular soil, taken from   \cite{Herle_Gudehus_Determination_of_parameters_of_a_hypoplastic_constitutive_model_from_properties_of_grain_assemblies}, is collected in Table \ref{tab:val_tip}. 

Finally, it is worth recalling that the SH constitutive law is a classical state-dependent model with the time arbitrarily scaled using a reference deformation ratio $\mathbf{D}$. The general constitutive law is in fact homogeneous and of first order with respect to $\mathbf{D}$, hence:

\begin{equation}\label{eq:lambaD1}
\accentset{\circ}{\mathbf{T}}(\mathbf{T},\lambda\mathbf{D},e)=\lambda\,\accentset{\circ}{\mathbf{T}}(\mathbf{T},\mathbf{D},e) \quad \textnormal{for} \quad \lambda>0 
\end{equation}

\begin{table*}[hbt!]
 \centering
  \caption{Parameters ($\mathbf{P}$) of the hypoplastic model for various granular soils \cite{Herle_Gudehus_Determination_of_parameters_of_a_hypoplastic_constitutive_model_from_properties_of_grain_assemblies}.}
    \begin{tabular}{llrrrrrrrr}
    \toprule
    \multicolumn{1}{c}{Soil} & \multicolumn{1}{c}{Tipe} & \multicolumn{1}{c}{$\varphi_c$ $(^\circ)$} & \multicolumn{1}{c}{$h_s$ $(MPa)$} &  \multicolumn{1}{c}{$n$ $(-)$}& \multicolumn{1}{c}{$e_{d0}$ $(-)$}  & \multicolumn{1}{c}{$e_{c0}$   $(-)$} & \multicolumn{1}{c}{$e_{i0}$  $(-)$}   & \multicolumn{1}{c}{$\alpha$ $(-)$}  & \multicolumn{1}{c}{$\beta$ $(-)$}  \\
    \midrule
    Hochstetten & gravel & 36    & 32000 & 0.18  & 0.26  & 0.45  & 0.50   & 0.10   & 1.9 \\
    Hochstetten & sand   & 33    & 1500  & 0.28  & 0.55  & 0.95  & 1.05   & 0.25  & 1.0 \\
    Hostun      & sand   & 31    & 1000  & 0.29  & 0.61  & 0.96  & 1.09   & 0.13  & 2.0 \\
    Karlsruhe   & sand   & 30    & 5800  & 0.28  & 0.53  & 0.84  & 1.00   & 0.13  & 1.0 \\
    Lausitz     & sand   & 33    & 1600  & 0.19  & 0.44  & 0.85  & 1.00   & 0.25  & 1.0 \\
    Toyoura     & sand   & 30    & 2600  & 0.27  & 0.61  & 0.98  & 1.10   & 0.18  & 1.1 \\
    Zbraslav    & sand   & 31    & 5700  & 0.25  & 0.52  & 0.82  & 0.95   & 0.13  & 1.0 \\
    \bottomrule
    \end{tabular}%
  \label{tab:val_tip}%
\end{table*}%

\subsection{Axisymmetric Conditions}

Following Herle \& Gudehus \cite{Herle_Gudehus_Determination_of_parameters_of_a_hypoplastic_constitutive_model_from_properties_of_grain_assemblies}, the assumption of axisymmetry of the tensor equation \eqref{eq:Wolffersdorff} simplifies both the principal (axial) stress $T_1$ and the second (radial) stress $T_2$. The stress and the rate of deformation tensors reduce to

\begin{equation}\label{eq:axisTD}
\mathbf{T}= \begin{bmatrix}T_1 & 0 &0 \\ 0 & T_2& 0 \\ 0&0&T_2 \end{bmatrix}  \quad  \mathbf{D}= \begin{bmatrix}D_1 & 0 &0 \\0 & D_2&0\\0&0&D_2 \end{bmatrix}\,.
\end{equation}

In the case of $\mathbf{W}=0$, the objective stress tensor reduces to the Cauchy effective stress $\accentset{\circ}{\mathbf{T}}=\dot{T}$ (see eq. \ref{eq:oT}) while introducing \eqref{eq:axisTD} in the nonlinear function $\mathbf{F}$ from \eqref{Eq1} yields:

\begin{multline}
\dot{T}_{1}=f_s \frac{(T_{1}+2T_{2})^2}{T_{1}^2+2T_{2}^2} \cdot   \Biggr[ D_1+a^2\biggl(\frac{T_{1}D_1+2T_{2}D_2}{(T_{1}+2T_{2})^2}\biggr)T_{1}+ \\ + f_d\frac{a}{3}\biggl(\frac{5T_{1}-2T_{2}}{T_{s1}+2T_{2}}\biggr)\sqrt{D_1^2+2D_2^2}\Biggl]\,,
\label{eq:T1}
\end{multline}

\begin{multline}
\dot{T}_2=f_s\frac{(T_1+2T_2)^2}{T_1^2+2T_2^2} \Biggr[ D_2+a^2\biggl(\frac{T_{1}D_1+2T_{2}D_2}{(T_{1}+2T_{2})^2}\biggr)T_2+\\ + f_d\frac{a}{3}\biggl(\frac{4T_{2}-T_{1}}{T_{1}+2T_{2}}\biggr)\sqrt{D_1^2+2D_2^2}\Biggl]\,.
\label{eq:T2}
\end{multline}

The response of the soil to the eodometric and triaxial drained tests can obtained by integrating in time the dynamical system in eq.s \eqref{eq:T1}-\eqref{eq:T2}, together with the continuity equation \eqref{eq:e_dot}.

\section{Calibration Methodology}\label{Tre}

The proposed calibration procedures combines a stochastic optimizer with a numerical solver of Ordinary Differential Equations (ODE). 

For every set of parameters $\mathbf{P}$, the ODE solver integrates the hypoplasticity model in \eqref{Eq1} to compute the soil response to a set of tests; the optimizer compares the obtained curves with a set of experimental data points and updates the parameters $\mathbf{P}$ until a maximum number of iteration is reached. All the function used in the calibration algorithm are developed in Python, using the Numpy library \url{https://numpy.org}. Both for the matrix operations and the random number generation \cite{Numpy}.

The procedure for integrating the hypoplasticity model is described in Section \ref{TrepUno} while the stochastic optimization strategy is described in Section \ref{TrepDue}. Section \ref{subsec_serarc} reports a note on the search space definition.

\subsection{Integration Procedure}\label{TrepUno}

Assuming that the eodometric and the triaxial drain test simulated in the calibration are in axisymmetric conditions, the hypoplasticity model simplifies to \eqref{eq:T1}, \eqref{eq:T2} and \eqref{eq:e_dot}.

The integration of this dynamical system is carried out using the simple explicit Euler scheme. This allows for keeping the computational cost of each integration to a minimum, minimizing the number of function evaluations. Moreover, this formulation allows for an easy check of the solution admissibility ($\Tr(\mathbf{T})<0$ and $e_d\leq e \leq e_i$) at every time step. Defining $\mathbf{X}^{k}=[T_1^{k}, T_2^{k}, e^{k}]^T$ the state vector of the ODE system, the time integration scheme reads

\begin{equation}
\label{Euler}
 \mathbf{X}^{k+1}=\mathbf{F}_{\mathbf{G}}(\mathbf{X}^{k},\mathbf{P})\,\Delta t +\mathbf{X}^{k}\,,
\end{equation} where $\mathbf{F}_{\mathbf{G}}$ here includes both $\mathbf{F}$ and $g$ in \eqref{Eq1}.

Because of the linear and homogeneous relation in \eqref{eq:lambaD1}, it is possible to fix an arbitrary reference $D_1=-1$ and compute the integration time $t_f$ from the maximal deformation obtained at the end of each test. For the oedometric test, the integration time is 

\begin{equation}
    t_f =  -\ln \biggr( 1- \frac{e_{0}-e_{fin}}{e_{0}+1}\biggl)\,,
\end{equation} while for the triaxial compression test is 

\begin{equation}
  t_f = \varepsilon_{fin}  \,.
\end{equation}
Two exemplary results from these two tests are shown in Figures \ref{fig:schema_Edo} and \ref{fig:schema_TxD} in which the $e_0,\,e_{fin},\,\varepsilon_{fin}$ are indicated.

For the sake of completeness, the deformation along the first principal component remains indicated as $D_1$, although all the calculations presented in this work implies $D_1=-1$. 

Once the integration time $[0, \,t_f]$ is defined, the time step is computed as $\Delta t=t_f / n_{Step}$, where the number of time steps is fixed to $n_{Step}=100$. The function $\mathbf{F}(\mathbf{X},\mathbf{P})$ in \eqref{Euler} differs in the two tests, as detailed in the following subsections.

\subsubsection{Oedometer Compressive test}\label{par:Edo}

The oedometer test consists in measuring the vertical displacement of a sample subject to vertical compression and having lateral expansion prevented. The sample for this test must be as loose as possible. A compacted dense sample, which has undergone load cycles with more than one reversal point, would in fact be difficult to model with the SH \cite{Wolffersdorff_A_hypoplastic_for_granular_material_with_a_predefined_limit_state_surface}. The procedure to correctly prepare the sample for calibrating the SH model is discussed in \cite{Kolymbas__Soft_Oedometer}.

\begin{figure}[h]
\centering
\includegraphics[width=0.45\textwidth]{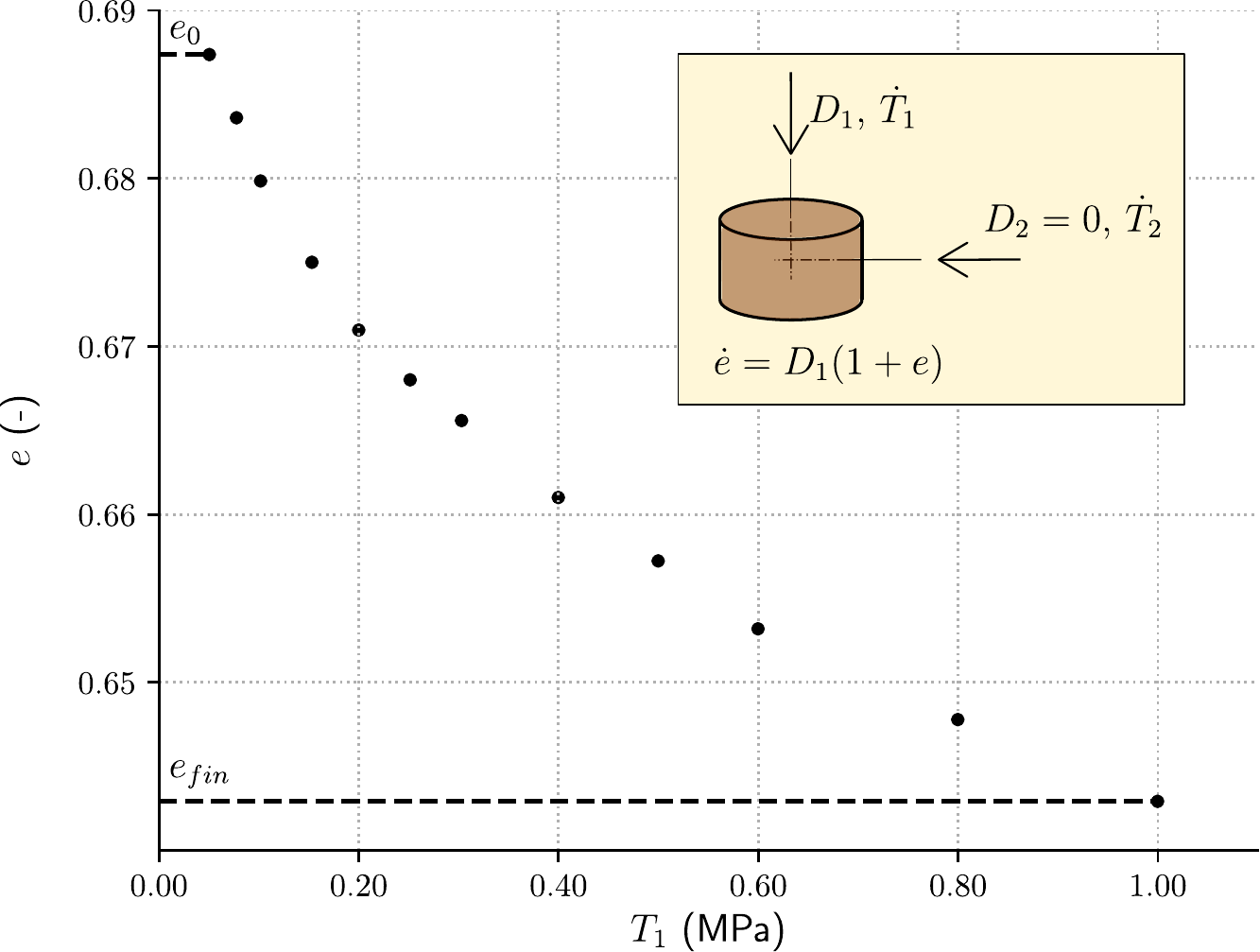}
\caption{Schematic of the eodometric test and exemplary set of points obtained from a tests.}
\label{fig:schema_Edo}
\end{figure}

A schematic of the test, recalling the main parameters involved is shown in Figure \ref{fig:schema_Edo} together with a sample set of experimental data. The results of this tests are usually collected in the plane ($e$, $-T_1$) for sand and ($e$, $\log(-T_1)$) for clay.

The boundary condition for the eodometric test is $D_2=0$. Imposing this to the set of equations \eqref{eq:T1},\eqref{eq:T2} and \eqref{eq:e_dot}, the system \eqref{Eq1} in matrix form reduces to:

\begin{multline}
\begin{bmatrix}
       \dot{T}_1\\[0.3em]
       \dot{T}_2\\[0.3em]
       \dot{e}\\
      \end{bmatrix} \quad  = f_s \begin{bmatrix}
       L_{11} & L_{12}&0\\[0.3em]
       L_{21} & L_{22}&0\\[0.3em]
       0 & 0&1+e\end{bmatrix}\,\begin{bmatrix}
       D_1\\[0.3em]
       0\\[0.3em]
       D_1\\
      \end{bmatrix}\,+
      \\  +f_s\,f_d \,
       \begin{bmatrix}
       N_1 \\[0.3em]
       N_2 \\[0.3em]
       0 \end{bmatrix} \sqrt{D_1^2}\,,
\label{eq:sistema_Edo}
\end{multline} where

\begin{align}
&L_{11}	= \Tr(\mathbf{T})^2/\Tr(\mathbf{T}^2)\cdot(1+a^2T_1^2/\Tr(\mathbf{T})^2)\\
&L_{12}	= 2a^2T_1T_2/\Tr(\mathbf{T}^2)\\
&L_{21}	= a^2T_1T_2/\Tr(\mathbf{T}^2)\\
&L_{22}	= 2a^2T_1T_2/\Tr(\mathbf{T}^2)\cdot(1+a^2T_2^2/tr(\mathbf{T})^2)\\
&N_{11}	= \Tr(\mathbf{T})/\Tr(\mathbf{T}^2)\cdot a/3(5T_1-2T_2)\\
&N_{22}	= \Tr(\mathbf{T})/\Tr(\mathbf{T}^2)\cdot a/3(4T_2-T_1) 
\end{align}



\subsubsection{Triaxial Compression Test}\label{par:Txd}
The triaxial compression test is performed on a drained and saturated sample, consolidated at a prescribed pressure \cite{Kolymbas_Recent_Results_of_Triaxial_Tests_with_Granular_Materials}. During the axial compression, the radial pressure is kept constant. The tests are performed at a controlled deformation rate $\varepsilon_a$: 

\begin{equation}
    \varepsilon_a=-\int_0^{t_f} D_1 \,dt=t_f\,.
\end{equation} 

The test returns the volume change $\varepsilon_v$

\begin{equation}
    \varepsilon_v=-\int_0^{t_f} \Tr(\mathbf{D}) \,dt=\frac{e-e_{0}}{1+e_{0}},
\end{equation} and the deviatoric stress 

\begin{equation}
    q=T_2-T_1\,.
\end{equation} 

\begin{figure}[!htb]
  \centering
  \subcaptionbox{Tiaxial deviatoric plane}[1\linewidth][c]{%
    \includegraphics[width=0.44\textwidth]{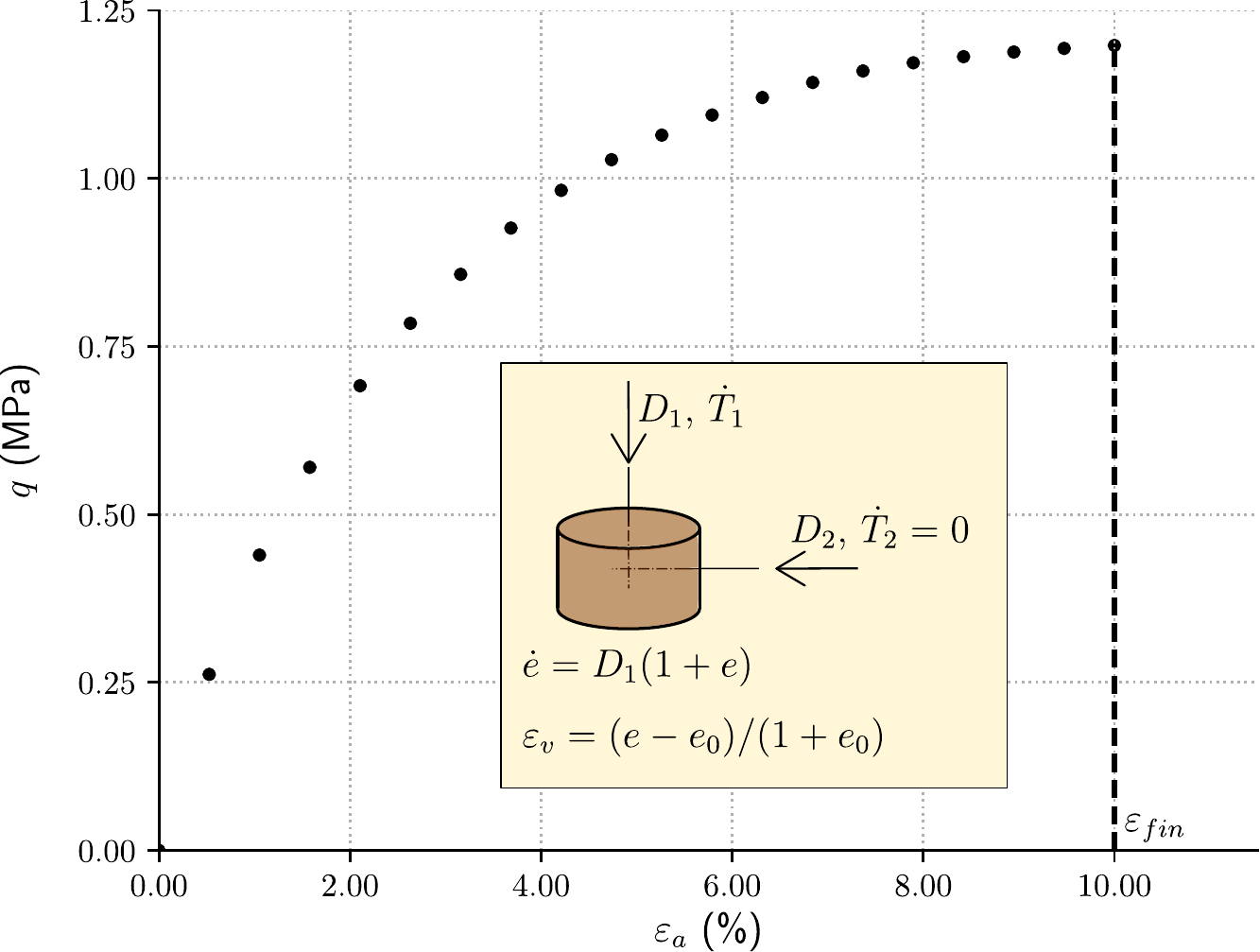}}\\
  \subcaptionbox{Triaxial volumetric plane}[1\linewidth][c]{%
    \includegraphics[width=0.44\textwidth]{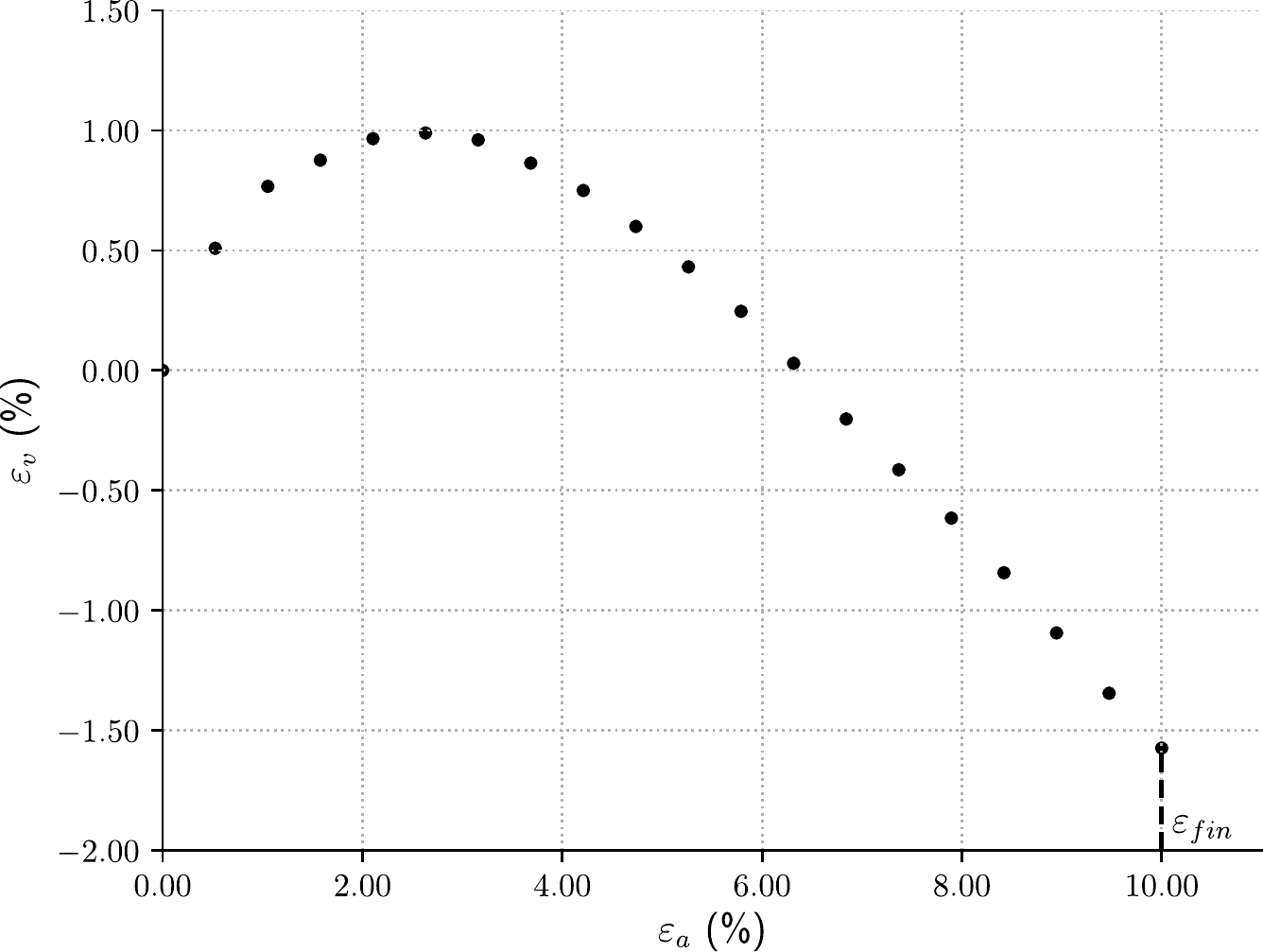}}
  \caption{Scheme recalling the main parameters of the triaxial test and example of resulting experimental points in the ($q,\varepsilon_a$, top) and the ($\varepsilon_v,\varepsilon_a$, bottom) planes.}
\label{fig:schema_TxD}
\end{figure}

A sketch of the test, with the relevant parameters and an example set of results are shown in Figure \ref{fig:schema_TxD}. The results of this test are usually given in triaxial deviatoric plane ($\varepsilon_a,\,q$) and triaxial volumetric plane ($\varepsilon_a,\,\varepsilon_v$).

The boundary condition for the triaxial drained test is $\dot{T}_2=0$. Therefore, the set \eqref{eq:T1}, \eqref{eq:T2} and \eqref{eq:e_dot} in matrix form becomes:
 
\begin{multline}
\begin{bmatrix}
       \dot{T}_1\\[0.3em]
       0\\[0.3em]
       \dot{e}\\
      \end{bmatrix} \quad  = f_s \begin{bmatrix}
       L_{11} & L_{12}&0\\[0.3em]
       L_{21} & L_{22}&0\\[0.3em]
       0 & 0&1+e\end{bmatrix}\,\begin{bmatrix}
       D_1\\[0.3em]
       D_2\\[0.3em]
       D_1+2\,D_2\\
      \end{bmatrix}\,+
      \\  +f_s\,f_d \,
       \begin{bmatrix}
       N_1 \\[0.3em]
       N_2 \\[0.3em]
       0 \end{bmatrix} \sqrt{D_1^2+2\,D_2^2}\,.
\label{eq:sistema_TxD}
\end{multline}

This problem is mixed since unknowns are both on RHS ($\dot{T_1}$ and $e$) and the LHS ($D_2$) \cite{Nova_Controllability_of_the_incremental,Imposimato_Uni,Kolymbas_Introduction_to_Hypoplasticity}. Following Nemunis \cite{Niemunis_Extended_hypoplastic_models_for_soils}, the solution strategy consists in obtaining an expression for $D_2$ from the second equation, solve the resulting quadratic in terms of the norm $x=\sqrt{D_1^2+2D_2^2}$, and finally consider only the solution with $x>0$. The second equation yields:

\begin{equation}\label{eq:TxD_D_2}
D_2=-\frac{f_d\,N_2\,x + L_{21}\,D_1}{L_{22}}  
\end{equation} and the introduction of $x$ gives:

\begin{equation}
\label{SOL}
x^2-2\,\frac{(-N_s\,f_d\,x-L_{11}\,D_1)^2}{L_{22}^2}-D_1^2 = 0 \,. 
\end{equation}

The solution of the resulting quadratic are:

\begin{multline}\label{eq:XI}
x_I=-\frac{L_{22}\sqrt{-2\,f_d^2\,N_2\,D_1^2+L_{22}^2\,D_1^2+2\,D_1^2\,L_{21}}+}{2\,f_d\,^2N_2^2-L_{22}} + \\  + \frac{2\,f_d\,L_{21}\,N_2\,D_1}{2\,f_d\,^2N_2^2-L_{22}}\,,
\end{multline}

\begin{multline}\label{eq:XII}
x_{II}=\frac{L_{22}\sqrt{-2\,f_d^2\,N_2\,D_1^2+L_{22}^2\,D_1^2+2\,D_1^2\,L_{21}}}{2\,f_d\,^2N_2^2-L_{22}} + \\ - \frac{2\,f_d\,L_{21}\,N_2\,D_1}{2\,f_d\,^2N_2^2-L_{22}}\,.
\end{multline}

The positive solution among $x_{I}$ e $x_{II}$ is used to compute $D_2$ from \eqref{eq:TxD_D_2} and finally advance the system \eqref{eq:sistema_TxD}. The uniqueness of the solution of the hypoplasticity problem depends on the existence of a single positive solution of \eqref{SOL}. Therefore, if multiple or no positive solutions exist, the proposed algorithm excludes the corresponding set of parameters.

\subsection{Cost Function Definition}\label{TrepDue}

The cost function driving the optimization of the parameters $\mathbf{P}$ is built by accounting for the discrepancy between the numerical predictions and the set of measurements (see also \cite{ZhenYu_Optimization_techniques_for_identifying_soil_parameters_in_geotechnical_engineering}). The formulation of the cost function must account for two important aspects. Firstly, the parameters are not entirely independent; secondly, the weight of the data in input should be unit-independent and have a comparable weight in the optimization, despite their largely different span (for example $e \in[0,4 - 1]$ while $q \in[0 - 1.6]$MPa).

Concerning the parameter independence, the optimal search must be constrained within the contractive/dilative domains of interest (cf. Figure \ref{fig:zone_di_comportamento}). This reduces the set $\mathbf{P}$ from eight to six parameters: $e_{d0}$ and $e_{i0}$ are chosen to preserve the ratios $\lambda_d$ and $\lambda_i$. Following Herle \& Gudehus  \cite{Herle_Gudehus_Determination_of_parameters_of_a_hypoplastic_constitutive_model_from_properties_of_grain_assemblies}, these ratios are taken in the range $\lambda_d=0.52 \div 0.65 $ and $\lambda_i\cong 1,2$. Therefore, the optimizer acts only on $\mathbf{P}^*$.

\begin{equation}
	\mathbf{P}^*=\{e_{c0},  h_s ,  \phi ,  n ,  \alpha ,  \beta \}\,.
\end{equation}

Concerning the weight of the data in the optimization, the experimental points obtained in the oedometric and the triaxial tests are scaled in the dimensionless planes ($\widehat{\varepsilon}_E  ,\widehat{T}_1$), ($\widehat{\varepsilon}_a , \widehat{q}$) and ($\widehat{\varepsilon}_a ,\widehat{\varepsilon}_v $). This new scaled set reads:

\begin{alignat}{1}
&\widehat{\varepsilon}_a =\varepsilon_a/\varepsilon_{fin}\\
&\widehat{\varepsilon}_v =\varepsilon_v/max(\varepsilon_v)\\
&\widehat{q}\,\, = q/max(q)\\
&\widehat{T}_1 = -T_1/min(-T_1)\\
&\widehat{\varepsilon}_E =-\ln\Bigr(1-\frac{e_0-e}{e_0+1}\Bigl)/\ln\Bigr(1-\frac{e_0-e_{fin}}{e_0+1}\Bigl) \,.
\end{alignat}

This scaling maps the experimental data onto curves that start from the origin ($0,0$) and end at ($1,1$).
 
\begin{figure}[hbt]
  \centering
    \includegraphics[width=0.45\textwidth]{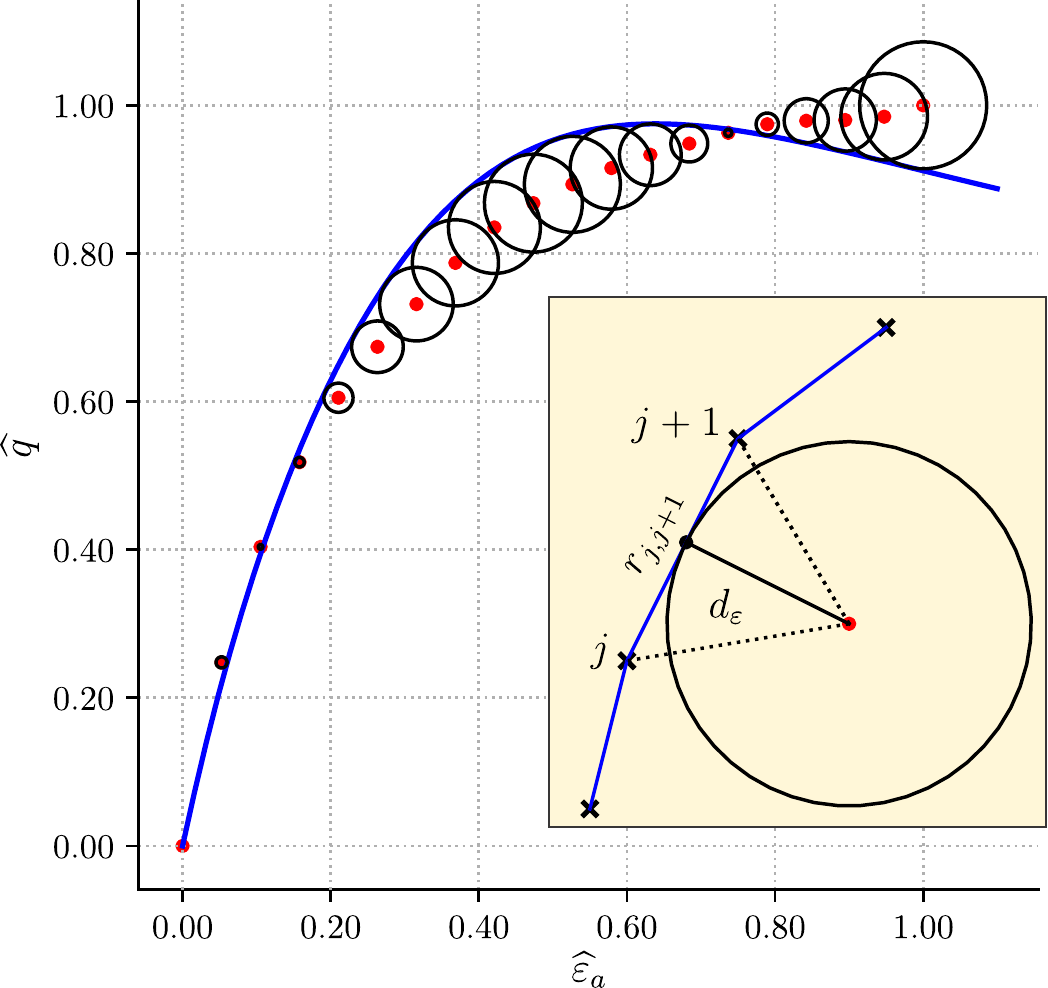}
\caption{Definition of Fr\'{e}chet distance between the experimental points and the prediction of the numerical model.}
\label{Example_NORMA}
\end{figure}

For a given set of parameters $\mathbf{P}$, the result of the numerical integration yields the time evolution of the solution vector $\mathbf{X}_{k}$ on a uniform temporal grid. The deviation between the set of experimental points and the model prediction is evaluated in terms of root mean square of the Fr\'{e}chet distance. This measurement of curve similarity has been already used in various applications and provides a measurement which is invariant to the axis orientation \cite{Jekel2018}. For a generic plane $(x,y)$, given a set of $M$ experimental points $(x_k,y_k)$ with $k\in[1,M]$ and a set of $N$ numerical predictions $(x_j,y_j)$ with $j\in[1,N]$, the discrete Fr\'{e}chet distance is a vector $D_{\mathcal{F}}$ of size $min(M,N)=M$ with entries

\begin{equation}
D_{\mathcal{F}}(k)=\min_{\forall j\leq N} \Bigl\{d_{\mathcal{E}}\bigl(k,r_{j,j+1} \bigr) \Bigr\}
\end{equation} where $d_{\mathcal{E}}(k,r_{j,j+1})$ is the distance between the experimental point $k$ and the segment line $r_{j,j+1}$ connecting two consecutive experimental points. These distances $d_{\mathcal{E}}$ are shown in the $\widehat{q}-\widehat{\varepsilon}_a$ plane in Figure \ref{Example_NORMA}. The Fr\'{e}chet distance is indicated with black circles centered on each data point, in red. In the figure, a sub-panel further describes the distance calculation. The deviation between experimental points and numerical prediction is finally computed as $\delta=||D_{\mathcal{F}}||_2$. 

This calculation is performed for the all curves produced by the tests, each providing a measurements of discrepancy. The final cost function is then  

\begin{equation}
\label{eq:Sco}
\mathcal{C} (\mathbf{P})= w_1 \,\delta_1(\mathbf{P}) + w_2 \,\delta_2(\mathbf{P})+ w_3 \,\delta_3(\mathbf{P}) 
\end{equation} where $w_{1,2,3}$ are the weights setting the relative importance of each plane. The cost function in \eqref{eq:Sco} can be easily extended to include results from other tests, if these are available. Moreover, while this work presents a single-objective optimization, multiple objectives can be implemented via multiple cost functions. For example, one could consider each of the contribution in \eqref{eq:Sco} as a different cost function, and seek the best compromise (Pareto front) among the different objectives. 

\subsection{Genetic Algorithm Optimizer}
Like many other population-based stochastic optimizers, the Genetic Algorithm (GA) starts from an initial population of possible solutions --in this work the model parameters $\mathbf{P}^*$-- and generates new sets by applying statistical operators. In the GA, these operators are designed to mimic the Darwinian theory of survival of the fittest. Borrowing from Biology, the GA terminology refers to each of the possible solutions as \emph{individual} and the statistical operations are referred to as \emph{elitism}, \emph{mutation}, \emph{selection} and \emph{cross-over}.

As shown in the pseudo-code listed in the Algorithm \ref{alg:Genetic_main}, the genetic algorithm involves three procedures. The first one is \textsc{Init.Pop}, to \emph{initialize the population}; the second is \textsc{Eval.Pop}, to \emph{evaluate the population}; the third is \textsc{Update.Pop}, to \emph{update the population}. A more detailed listing of each of these procedures is provided in the algorithm \ref{alg:Genetic_Init.Pop}, \ref{alg:Genetic_Eval.Pop} and \ref{alg:Genetic_Update.Pop}. 

\begin{algorithm}
  \begin{algorithmic}
\State main():
\State \textsc{Init.Pop}($N_{i}$,$\mathbf{P}_{min}$,$\mathbf{P}_{max}$)$\rightarrow \mathbf{Pop}$
  \For {$iteration$ in (1,$N_I$)}
    \State \textsc{Eval.Pop}($\mathbf{Pop}$,$\lambda_i$,$\lambda_d$)$\rightarrow \mathbf{ID}$
    \State \textsc{Update.Pop}($\mathbf{Pop}$,$N_i$,$iteration$,$N_{I}$)$\rightarrow \mathbf{Pop}$
 \EndFor
\State \textsc{Eval.Pop}($\mathbf{Pop}$,$\mathbf{Edo}$,$\mathbf{Txd}$,$\mathbf{Edo}_0$,$\mathbf{Txd}_0$)

\State $\mathbf{P}^*$=$\mathbf{Pop}[\mathbf{ID}[1]]$ 
\State return:  $\mathbf{P}^*$ 
\State end
\caption{Optimization Algorithm--main}\label{alg:Genetic_main}
\end{algorithmic}
\end{algorithm}

A population is a matrix collecting all the individuals, one for each row. This matrix is indicated as $\mathbf{Pop}$. The size of $\mathbf{Pop}$ is ($N_i$,6), were $N_i$ is the number of individuals, and 6 is the size of constrained parameters $\mathbf{P}^*$. The population is initialized randomly within the search space bounded by vectors containing the lowest and the largest possible values of each parameter $\mathbf{P}^*_{min}$ and $\mathbf{P}^*_{max}$. These are introduced as user inputs.


As reported in the \textsc{Init.Pop} procedure, we initialize half of the population with a uniform distribution spanning the entire search space, the other half as a Gaussian distribution centered in the search space with a standard deviation equal to one-sixth of the range. 

\begin{algorithm}
  \begin{algorithmic}
\Procedure {Init.Pop}{$N_{i}$,$\mathbf{P}^*_{min}$,$\mathbf{P}^*_{max}$,$\lambda_i$,$\lambda_d$}
\State $N_{Gau}$=$0.5N_{i}$
\State $N_{Uni}$=$N_{i}-N_{Gau}$
 \For {$i$ in (1,6)}
    \State  $\mu[i]$   =($P^*_{max}[i]$-$P^*_{min}[i]$)/2
     \State $\sigma[i]$=($P^*_{max}[i]$-$P^*_{min}[i]$)/6
 \EndFor
 \State $\mathbf{P}_{U}$=random.uniform ($\mathbf{P}_{min}$,$\mathbf{P}_{max}$,$N_{Uni}$)
 \State $\mathbf{P}_{G}$=random.normal ($\mathbf{\mu}$,$\mathbf{\sigma}$,$N_{Gau}$)
 \State return:  $\mathbf{Pop}$ = $\mathbf{P}_{U}$ $\cup$ $\mathbf{P}_{G}$
\EndProcedure
\caption{Initialization of population.\\ \small{The functions random.uniform and random.normal are the ones available in Numpy \cite{Numpy}.}}\label{alg:Genetic_Init.Pop}
\end{algorithmic}
\end{algorithm}

Starting from the initial population, the \textsc{Eval.Pop} and the \textsc{Update.Pop} are executed in a loop until the maximum number of iterations $N_I$ is reached. The evaluation consists of computing the cost function of each set of parameters, i.e.,
of each individual. The cost associated with each individual is used by UPDATE.POP as a measurement of fitness and the population is ranked from the best (low cost) to the worst (high cost) candidate solution. This evaluation procedure is performed by the function EVAL.POP, which returns the index vector of the list of first placed individuals $\mathbf{ID}$.

\begin{algorithm}
  \begin{algorithmic}
\Procedure {Eval.Pop}{$\mathbf{Pop}$,$\lambda_i$,$\lambda_d$}
 \State $\mathbf{COST}$= C($\mathbf{Pop}$,$\lambda_i$,$\lambda_d$)
 \State $\mathbf{ID}$=$\mathbf{COST}$.argsort()
 \State return:  $\mathbf{ID}$
\EndProcedure 
\caption{Evaluate population}\label{alg:Genetic_Eval.Pop}
\end{algorithmic}
\end{algorithm}

The procedure \textsc{Update.Pop} update the population  combining \emph{elitism}, \emph{mutation}, \emph{selection} and \emph{cross-over}. \emph{Elitism} consists in advancing some the best individual to the next generation. The fraction of elite individuals is herein indicated with $n_E$ and the total number of elite individual passed to the next generation, $N_E=n_E\cdot N_i$, is taken form the $\mathbf{Pop}$ using the pointer form the first $N_E$ element of $\mathbf{ID}$ vector.

\emph{Mutation} is the fundamental operation that lets the GA explore the solution space: a percentage of the population at each iteration continues to be randomly chosen, in this work from a uniform random distribution in $\mathbf{P}^*_{min} \times \mathbf{P}^*_{max}$. The fraction of mutated individuals is indicated with $n_M$, and is computed as an exponentially decaying function of the iterations. This allows for balancing exploration and exploitation as the population convergences to its final distribution. 

The remaining $N_n=N_i (1-n_E-n_M)$ elements are generated from the best individual via \emph{selection} and \emph{cross-over}. \emph{Selection} is the operation that defines which of the individual is allowed to mate; \emph{cross-over} is the operation that defines how the information in mating individuals is combined to produce the new ones, referred to as the offspring. Following the rank weighting approach in \cite{Haupt2003}, the selection of individuals is performed using a set of random numbers. These are sampled from a triangular probability density function of the form:

\begin{equation}
\label{distri}
    p(n)=\frac{2(N_f-n)}{(N_f-1)^2},
\end{equation} rounded to the closest integer, where $N_f=n_f\,N_i$ is the number of individuals that is allowed to mate and $n\in[1,N_f]$ is the rank of the individual, namely the index in the sorted list $\mathbf{ID}$. This distribution implies that the fittest individuals ($n=1$) have a higher chance of mating while the last ($n=N_f$) has zero chances. 

Once the best $N_f$ individuals are identified, the \emph{cross-over} is generated by blending the features in the two parents as 

\begin{equation}
 \mathbf{P}_{new}=\theta_k \mathbf{P}_{n_1}+(1-\theta_k) \mathbf{P}_{n_2}     
\end{equation} where $n_1$ and $n_2$ are the indices of the two randomly chosen parents from the triangular distribution in \eqref{distri} and the vector $\theta_k$ selects a random number in the range $[0,1]$ for each of the $6$ entries in $\mathbf{P}_{n_1},\mathbf{P}_{n_2}$. The procedure \textsc{Update.Pop} returns a new population of individuals, characterized by an improved average cost. 

\begin{algorithm}
  \begin{algorithmic}
\Procedure {Update.Pop}{$\mathbf{Pop}$,$N_i$,$\mathbf{ID}$,$IT$,$N_{I}$}
 \State $n_E$=$0.01$, $n_f$=$0.50$, $\mu_0$=$0.5$, $\mu_{fin}$=$0.1$
 \State $N_E$=$n_E\cdot N_i$
 \State $\mathbf{P}_{eli}$=$\mathbf{Pop}[\mathbf{ID}[0:N_E]]$
 \State $n_M$=$\mu_0\cdot \exp[IT\,/N_{I}\log(\mu_{fin}/\mu_{0})] $
 \State $N_M$=$n_M\cdot N_i$
 \State $\mathbf{P}_{mut}$=random.uniform ($\mathbf{P}_{min}$,$\mathbf{P}_{max}$,$N_M$)
 \State $N_N$=$N_i\cdot(1-n_E-n_M)$ 
  \For {$i$ in (1,$N_N$)}
    \State $\mathbf{S}_{el}$=random.triangular($n_f\cdot N_N$,2)
    \State $\mathbf{\theta}$=random.uniform (0,1,6) 
    \State $\mathbf{P}_{n1}$=$\mathbf{Pop}$[$\mathbf{ID}$[$\mathbf{S}_{el}$[1]]
    \State $\mathbf{P}_{n2}$=$\mathbf{Pop}$[$\mathbf{ID}$[$\mathbf{S}_{el}$[2]]
    \State $\mathbf{P}_{new}$[i]=$\theta\cdot \mathbf{P}_{n1}+(\theta-1)\cdot \mathbf{P}_{n2}$
 \EndFor
 \State return:  $\mathbf{Pop}$ = $\mathbf{P}_{eli}$ $\cup$ $\mathbf{P}_{mut}$ $\cup$ $\mathbf{P}_{new}$
\EndProcedure 
\caption{Update the population.\\
\small{The functions random.uniform and random.triangular are the ones available in Numpy \cite{Numpy}.}}\label{alg:Genetic_Update.Pop}
\end{algorithmic}
\end{algorithm}

\subsection{A note on the search space}\label{subsec_serarc}

In the methodology proposed thus far, the model calibration is entirely entrusted to the Genetic Algorithm (GA). Indeed, the optimization can identify the correct parameters only if these are within the algorithm's search space. However, the proper definition of such search space requires experience and, in some cases, multiple trials. Increasing the search space increases the risks of encountering a local minimum and decreases the convergence performances of the optimization; decreasing the search space decreases the probability that the best set of parameters is included and hence reachable.

While it is not trivial to correctly identify the search space, it is generally easy to see if the chosen one is inappropriate: when this is too narrow, the population tends to clusters on its boundaries; when this is too large, a substantial variance between the solutions obtained in different trials is observed. It is thus essential to run the optimization several times and analyze the statistics of the identified parameters. This analysis is proposed in Section \ref{subpar:ConvergenzaMetodo}.  

It is good practice to build the calibration by using as much as possible well-known results from previous authors. In particular, some coefficients are more easily estimated than others. The coefficient $\varphi_c$, for example, can be obtained with usual procedures based on the Mohr plane with acceptable uncertainties, if the shear banding is prevented \cite{doi:10.1002/nag.338}. The parameters $n$ e $h_s$ can be estimated from the methods proposed in \cite{Kadlicek_Calibration_of_Hypoplastic_Models_for_Soils}. 
From the authors' experience, these can lead to estimations of $n$ with uncertainties in the range 10-20\%, while the uncertainty in the estimation of $h_s$ can reach up to 70\%. The remaining parameter can be estimated from the relations proposed in \cite{Herle_Gudehus_Determination_of_parameters_of_a_hypoplastic_constitutive_model_from_properties_of_grain_assemblies}.

\section{Results}\label{Quattro}

This section is organized in three subsections. In \ref{subpar:Validazione}, the numerical method to integrate presented in the section \ref{TrepUno} is validated using a free tool. In \ref{subpar:ConvergenzaMetodo}, the repeatably and the uncertainty of the calibration parameter is analyzed, along with a correlation analysis of the calibration parameters. Finally, \ref{subpar:ConvergenzaMetodo} compares the calibration results for the Hochstetten sand soil presented in \cite{Wolffersdorff_A_hypoplastic_for_granular_material_with_a_predefined_limit_state_surface} and \cite{Herle_Gudehus_Determination_of_parameters_of_a_hypoplastic_constitutive_model_from_properties_of_grain_assemblies}.

\subsection{Validation of the response curve}\label{subpar:Validazione}

The validation of the numerical model described in \ref{par:Edo} and \ref{par:Txd} was carried out using ExCaliber-Laboratory Test Simulation\footnote{see \url{https://soilmodels.com/excalibre-en/}}. This tool is developed by Prof. Ma\v{s}\'in and co-workers \cite{info_sol_model_projecr} and is powered by GEO5 FEM, a software by Fine Civil Engineering Software.

The hypoplastic parameters chosen for the validation are those proposed by von Wolffersdorff for the Hochstetten sand in \cite{Wolffersdorff_A_hypoplastic_for_granular_material_with_a_predefined_limit_state_surface}: $\varphi_c=33^{\circ}$, $h_s=10^6$ kPa, $n=0.25$, $e_{c0}=0.95$, $e_{d0}=0.55$, $e_{i0}=1.05$, $\alpha=0.25$ e $\beta=1.5$. 

\begin{figure}[h!]
  \centering
  \subcaptionbox{Eodometric plane}[1\linewidth][c]{
    \includegraphics[width=0.44\textwidth]{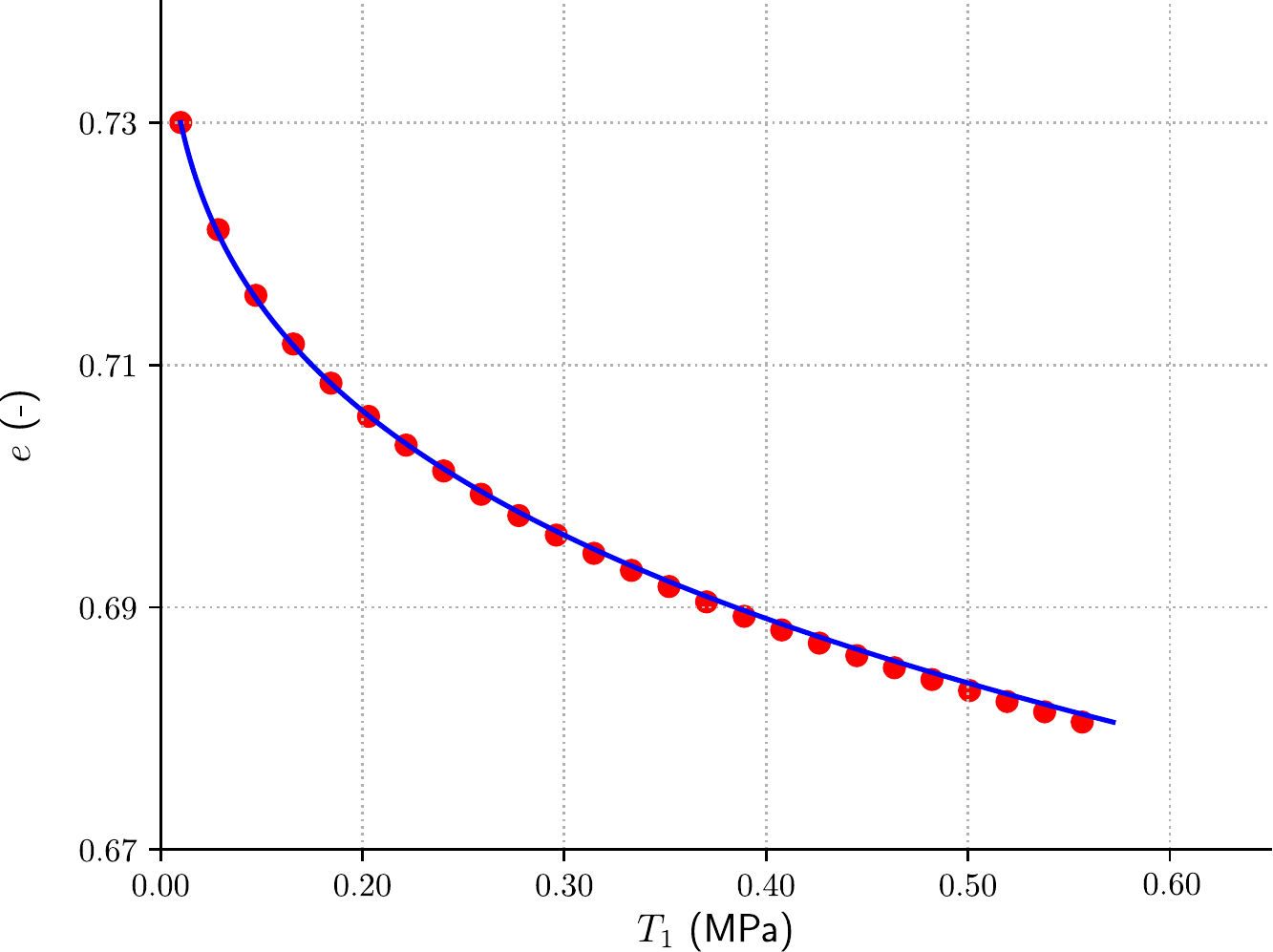}}\\
  \subcaptionbox{Tiaxial deviatoric plane}[1\linewidth][c]{%
    \includegraphics[width=0.44\textwidth]{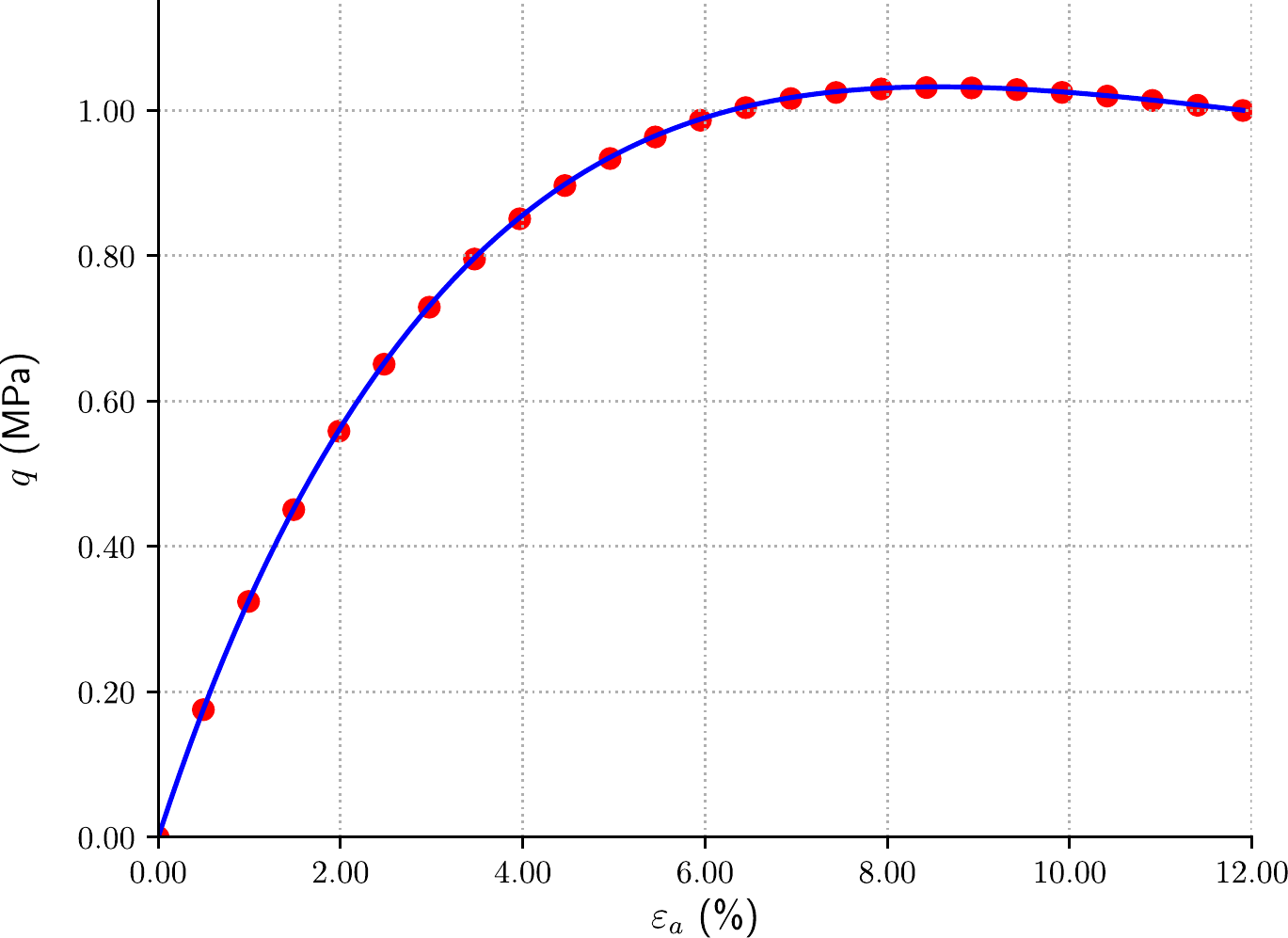}}\\
  \subcaptionbox{Triaxial volumetric plane}[1\linewidth][c]{%
    \includegraphics[width=0.44\textwidth]{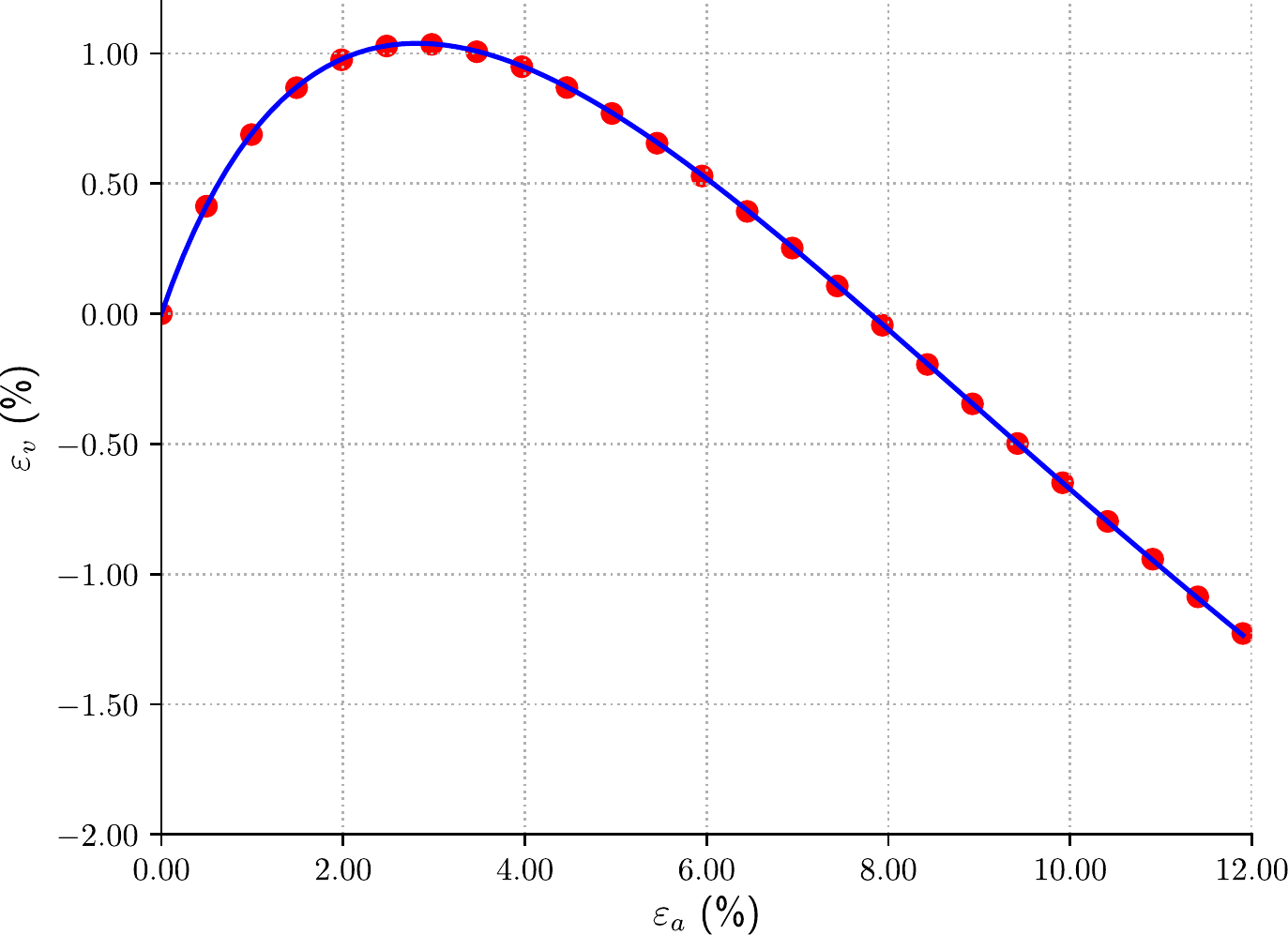}}
  \caption{Validation of the numerical model used in the GA calibration (blue continuous curve) with the ExCalibre-Laboratory Test software (red markers). The validation is performed on the Eodometric plane (a), the triaxial  deviatoric plane (b) an the triaxial volumetric plane (c).}
\label{fig:Validazione}
\end{figure}

The initial conditions are  $T_1=-300$kPa, $T_2=-300$ kPa and $e=0.660$ for the triaxial test and $T_1=-10$kPa, $T_2=-10$ kPa and $e=0.730$ for the oedometric test. The triaxial test goes up to the maximum deformation $\varepsilon_{fin}=0.11$  while the eodomeric test proceeds until the void ratio $e_{fin}=0.680$ is reached. The results for the three tests are shown in Figure \ref{fig:Validazione}. The curves are practically indistinguishable, hence validating the numerical procedure used by the proposed optimizer.

\subsection{Calibration Repeatability and Uncertainty}\label{subpar:ConvergenzaMetodo}

In order to validate the optimizer and analyze the uncertainty of the calibrated parameters, this section reports on the analysis of synthetic data.
The scope of these synthetic laboratory experiments, for which the exact set of model parameters is known, is threefold.

The first objective is to analyze how quickly and how well the optimizer converges to the final set of parameters. The second objective is to analyze the variance and hence the uncertainty of each parameter. It is worth highlighting that by \emph{uncertainty} we here refer to a measure of the parameter uniqueness. In other words, given a large set of converged solutions, all equally valid according to the cost function in \eqref{eq:Sco}, we reveal how sensitive the model is with respect to a given parameter. The third objective is to analyze the correlation between all the parameters and hence open possible avenues for a data-driven reduction of the calibration problem. 

The parameter chosen for the simulations in this section are $\varphi_c=34^{\circ}$, $h_s=3.8\cdot 10^6$ $kPa$, $n=0.30$, $e_{c0}=0.886$, $e_{d0}=0.531$, $e_{i0}=1.06$, $\alpha=0.144$ and $\beta=1.5$. These represent the \emph{exact} solution for the calibration procedure. A set of $M=15$ points is extracted from the numerical simulation of one eodometer test while $M=30$ points is extracted from three triaxial drained tests. These tests, one odometer test and three triaxial drained tests, provide the minimal requirement for the model calibration.

The initial conditions for these synthetic tests are shown in the Table \ref{tab:cond_contorno_val}.  The triaxial tests go on until the maximum deformation $\varepsilon_{fin}=0.20$ is reached, while the eodomeric tests continue until a void ratio $e_{fin}=0.720$. 

\begin{table}[hbt!]
 \centering
  \caption{Initial condition for the triaxial tests (TxD1,TxD2,TxD3 ) and the oedometer test (EDO1) for the synthetic test cases.}
    \begin{tabular}{lrrr}
    \toprule 
    \multirow{2}{*}{Test} & \multicolumn{1}{c}{$T_1$ } The soil& \multicolumn{1}{c}{$T_2$} & \multicolumn{1}{c}{$ e$ }  \\
   
    & \multicolumn{1}{c}{($kPa$)} & \multicolumn{1}{c}{($kPa$)} & \multicolumn{1}{c}{($-$)}  \\
    \midrule
    TxD1 & - 50.0    & - 50.0 &   0.524 \\
    TxD2 & -100.0    & -100.0 &   0.545 \\
    TxD3 & -200.0    & -200.0 &   0.588 \\
    EDO1 & -8.0     &  -4.0   &   0.784 \\
    \bottomrule
    \end{tabular}%
  \label{tab:cond_contorno_val}%
\end{table}%

We consider ratios $\lambda_d=0.60$ and $\lambda_i=1.20$  and the search interval indicated in table \ref{tab:Int_1}.

\begin{table}[hbt!]
 \centering
  \caption{Search space bounded vectors  $\mathbf{P}^*_{max}$and $\mathbf{P}^*_{min} $ }
    \begin{tabular}{lrrrrrr}
    \toprule 
    \multirow{2}{*}{ID}&\multicolumn{1}{l}{$\varphi$ } & \multicolumn{1}{c}{$h_s$} & \multicolumn{1}{c}{$n$} & \multicolumn{1}{c}{$e_{c0}$} & \multicolumn{1}{c}{ $\alpha$ } & \multicolumn{1}{c}{ $\beta$  }\\
     &\multicolumn{1}{l}{($^{\circ})$} & \multicolumn{1}{c}{ (GPa)} & \multicolumn{1}{c}{(-)} & \multicolumn{1}{c}{(-)} & \multicolumn{1}{c}{ (-) } & \multicolumn{1}{c}{ (-) }\\
    \midrule
    max &40 & 9.0 & 0.40 & 1.1 & 0.20 & 2.0 \\
    min &25 & 1.0 & 0.25 & 0.6 & 0.05 & 1.0 \\
    \bottomrule
    \end{tabular}%
  \label{tab:Int_1}%
\end{table}%

The parameters of the GA are set to $ N_{i} = 500 $ and $ N_I = 20 $, while the remaining ones are taken as the default in Algorithm \ref{alg:Genetic_Init.Pop} and \ref{alg:Genetic_Update.Pop}. The weights $ w_i $ $ i = 1,2,3 $ in the cost function \eqref{eq:Sco} are equal to unity, hence giving equal importance to the errors in each test.

\begin{figure}[t]
  \centering
  \subcaptionbox{}[0.9\linewidth][c]{%
    \includegraphics[width=0.43\textwidth]{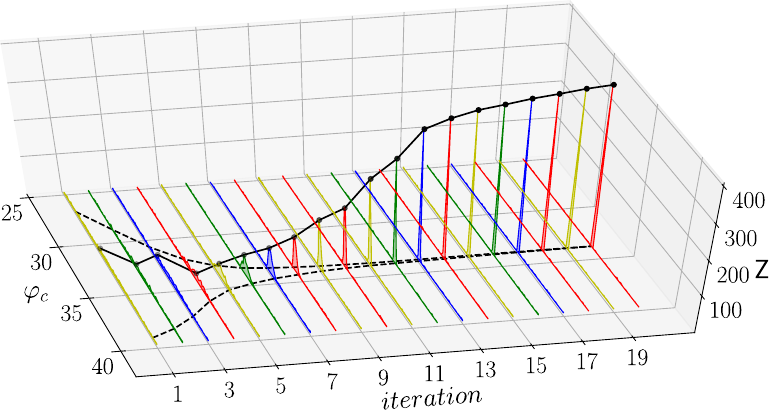}}\\
  \subcaptionbox{}[0.9\linewidth][c]{%
    \includegraphics[width=0.41\textwidth]{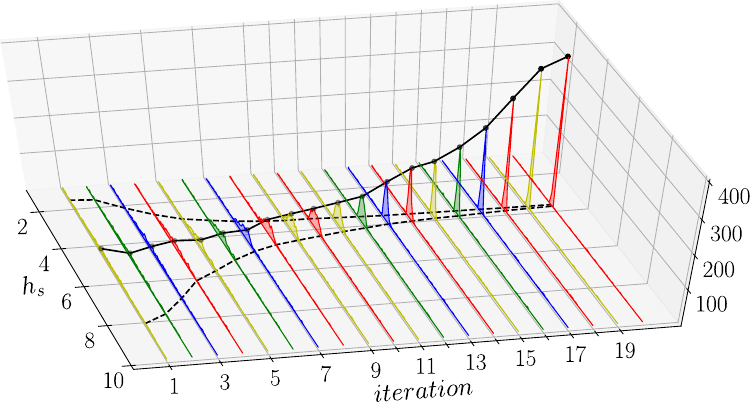}}
  \caption{Evolution of the distribution of $\varphi_c$ (a) and $h_s$ (b) over the iterations, showing the convergence of the population. The dashed lines mark the interval $ \mu \pm 2 \sigma $ at each iteration, where $\mu $ and $\sigma $ are the mean and the standard deviation of the population allowed to mate.}
\label{fig:3D_ist}
\end{figure}

To qualitatively analyze the convergence of the GA, we first focus on the evolution of the distribution of parameters during the iterative search. For the sake of compactness, we here focus on the histograms of the parameters $\phi_c$ and $h_s$, being the histograms of the others quite similar. The evolution of the population of these two parameters are shown in Figure \ref{fig:3D_ist}. For both, the initial population has a rather flat histogram, with a slightly larger concentration in the central area of the search space, as prescribed in the Algorithm \ref{alg:Genetic_Init.Pop}. Iteration by iteration, the distribution focuses on the result that minimizes the cost function, and the peak in the histograms grows accordingly. The narrowing of the population distribution is further highlighted by the dashed lines in Figure \ref{fig:3D_ist}; these lines mark the boundaries of the interval $ \mu \pm 2 \sigma $ at each iteration, where $\mu $ and $\sigma $ are the mean and the standard deviation of the population allowed to mate.

The rate of convergence largely depends on the sensitivity of the cost function to each parameter: in the figures shown, the parameter $\phi_c$ appears to have a more important impact, and hence its distribution converges faster than $h_s$. The reader should notice that even if the convergence is reached in both cases after about 15 iterations, a small portion of the histogram remains flat and outside the mean value. This is due to the small percentage of mutations that is maintained through the iterations to continue exploring the search space.

\begin{figure}[h]
\centering
\includegraphics[width=0.45\textwidth]{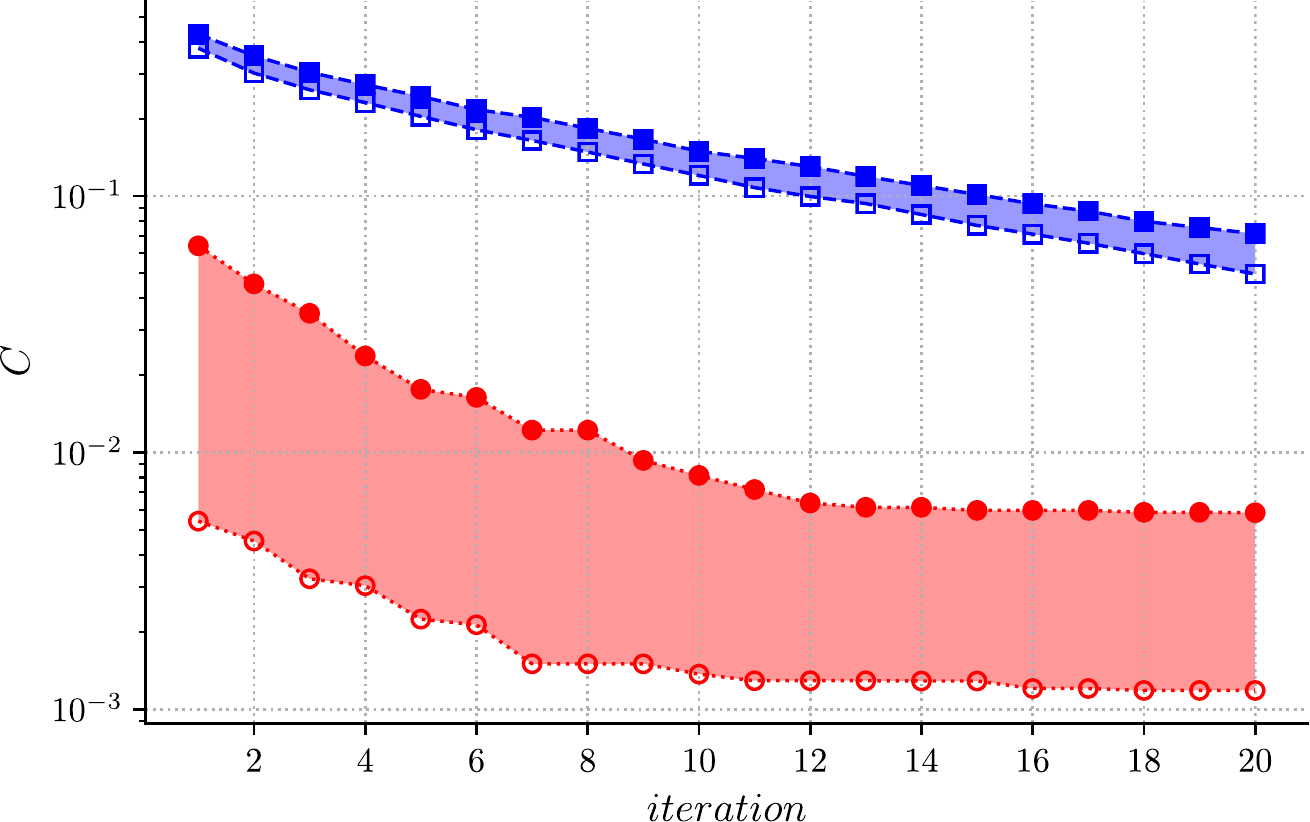}
\caption{Reduction of the cost function over the iterations. The red circles indicate the max values, the blue squares indicate the mean values. For both quantities, the plot shows the best-case (denoted with filled markers) and the worst-case (denoted with empty markers) over 1000 tests.}
\label{fig:Convergenza_errore}
\end{figure}

\begin{figure}[h!]
  \centering
  \subcaptionbox{Eodometric plane}[1\linewidth][c]{%
    \includegraphics[width=0.44\textwidth]{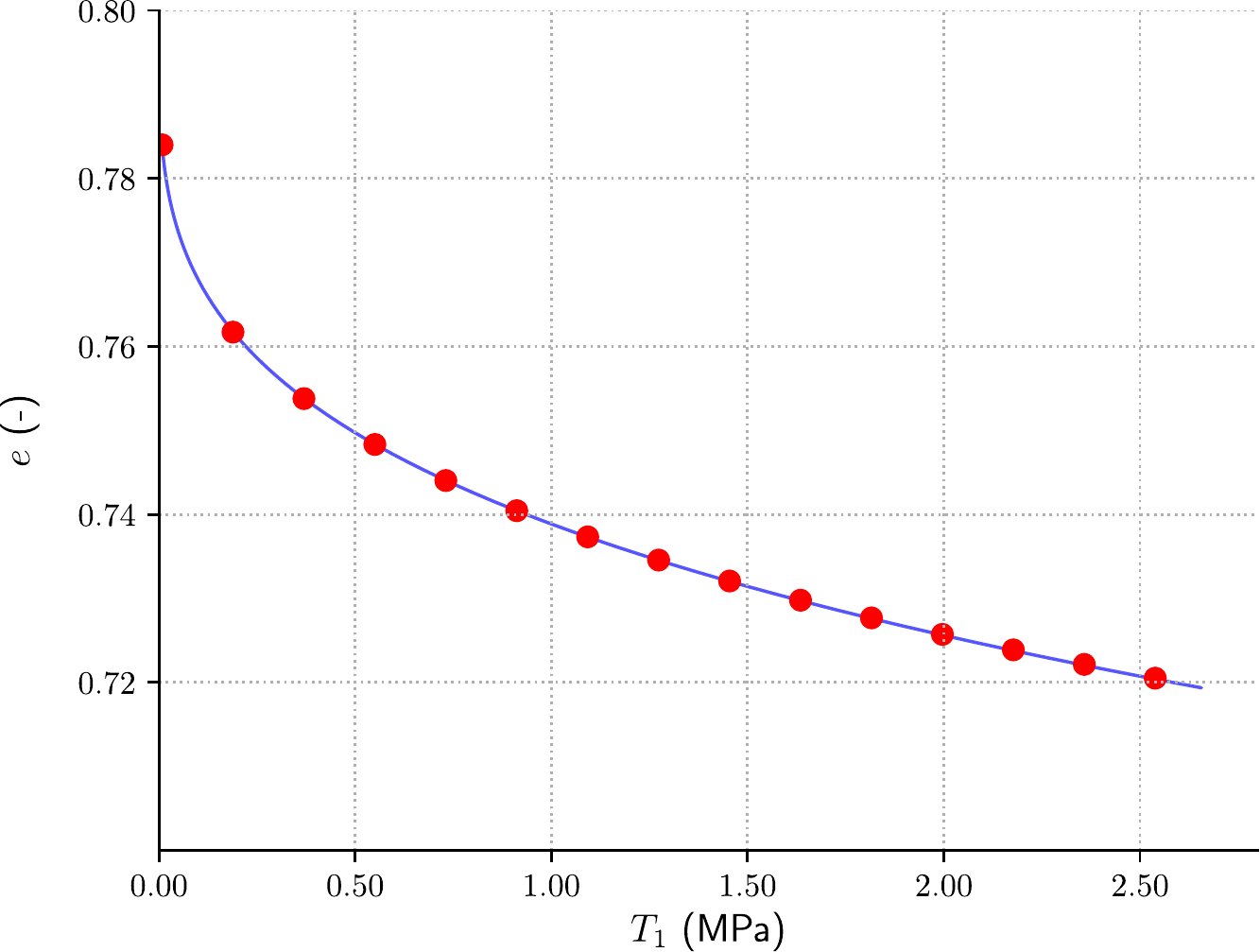}}\\
  \subcaptionbox{Tiaxial deviatoric plane}[1\linewidth][c]{%
    \includegraphics[width=0.44\textwidth]{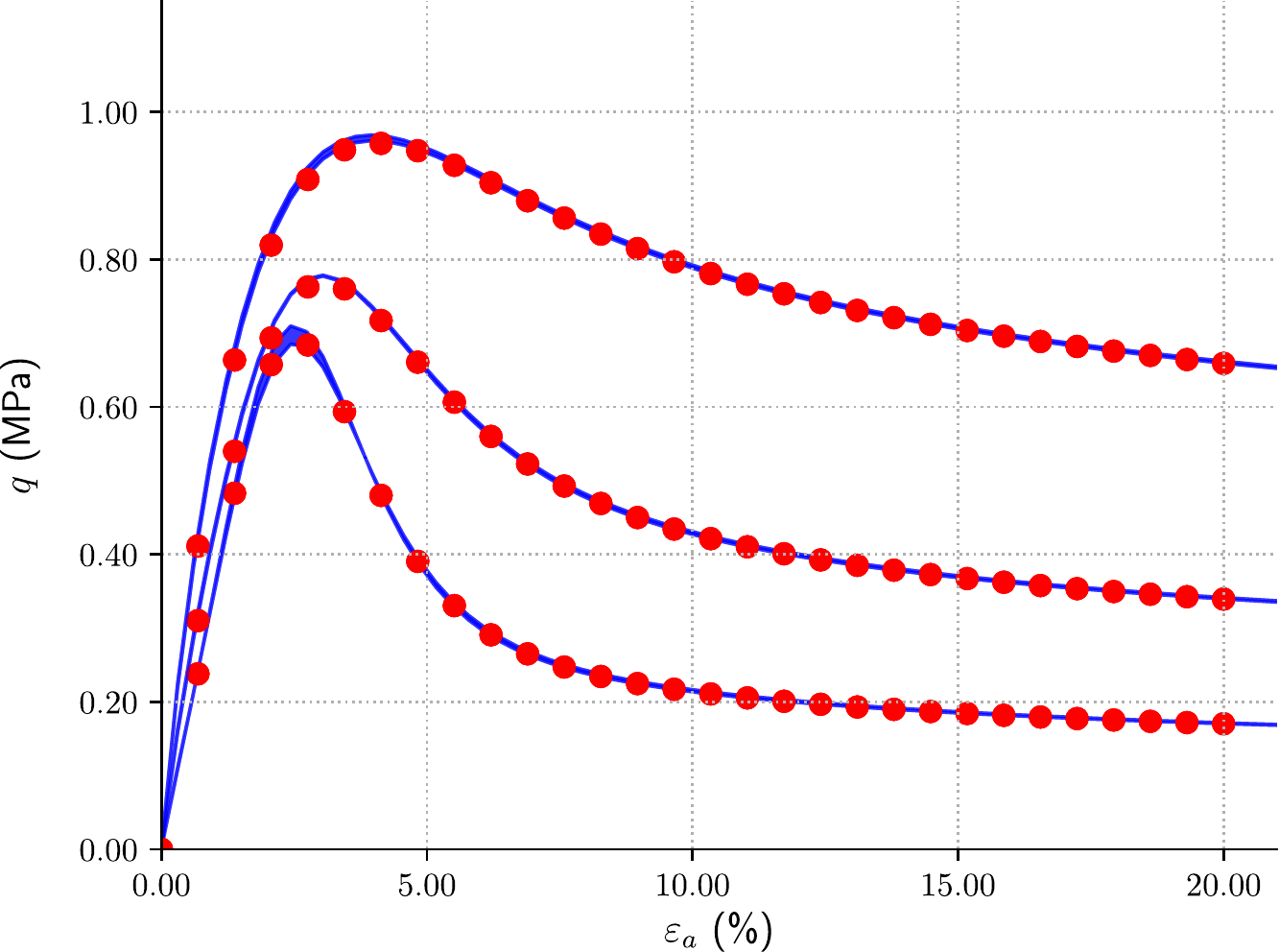}}\\
  \subcaptionbox{Triaxial volumetric plane}[1\linewidth][c]{%
    \includegraphics[width=0.44\textwidth]{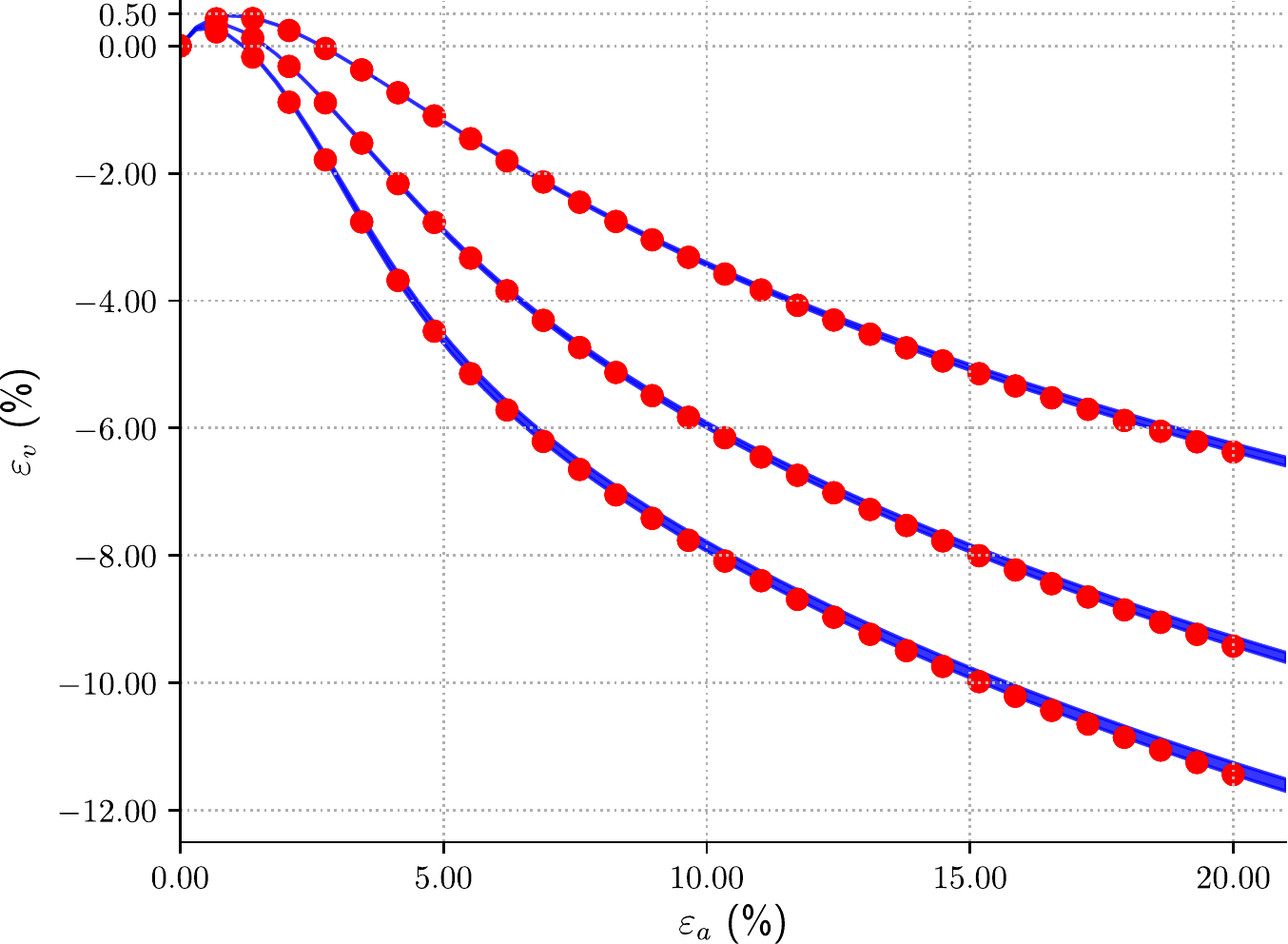}}
  \caption{Validation of the GA calibration on the Eodometric plane (a), the triaxial  $(\varepsilon_a,\,q)$  plane (b) and $(\varepsilon_a,\,\varepsilon_v)$ plane (c). The red circles indicate the synthetic experimental data. The blue lines showthe numerical prediction using the best and the worst set of parameters from the full set: as these are practically overlapping, the comparison appears as a thick line. }
\label{fig:Convergenza}
\end{figure}

To assess the convergence performance of the algorithm and the parameter uncertainties, the calibration is here repeated 1000 times. Figure \ref{fig:Convergenza_errore} collects the main results on the cost function evolution as a function of the iteration number. The plot shows the evolution of the mean error, indicated with blue square markers, and the minimal error, indicated with red circle markers. For each of these quantities, the upper curve refers to the worst possible result among the 1000 trials, while the lowest curve refers to the best result. As expected, the convergence is proven by a reduction of one to two orders of magnitudes in the cost function. To further highlight the optimization convergence, Figure \ref{fig:Convergenza} compares the experimental results with the prediction of the numerical solver using the best and the worst set of parameters obtained from the last iteration of all the trials. As the difference in the cost function varies from $C(\mathbf{P})=3\cdot 10^{-1}$ (worst case) to $C(\mathbf{P})=4\cdot 10^{-2}$ (best case), the difference in the prediction is unnoticeable.

It thus safe to conclude that the algorithm has converged, and setting the maximum number of iterations to $N_I=20$ ensures that both the best-case and the worst-case set lead to acceptable results. The Figure \ref{fig:Convergenza_errore} also shows that satisfactory convergence is reached after about ten iterations. However, despite the satisfying convergence, it is essential to notice that the final cost function is still three orders of magnitude larger than the cost function associated with the \emph{exact} (the introduced) solution, which leads to $C(\mathbf{P})=2\cdot 10^{-5}$.

Because of this apparently irrelevant difference, the obtained set of parameters does not coincide with the exact one. Moreover, the statistics of the parameters obtained in all the trials lead to a non-negligible variance, which can be associated to the parameter uncertainty. The main statistical results for each of the parameters, obtained over all the tests, are collected in Table \ref{tab:varianza-par}. In particular, the table collects the mean result, the standard deviation normalized by the mean, the minimum and the maximum values. While for most parameters the ratios $\sigma/\mu$ are below $3\%$, the normalized standard deviation $\sigma/\mu$ for $h_s$ reaches up to $7\%$. This implies that this parameter is overall less important than the others and its precise estimation is of comparatively lower importance.

\begin{table}[hbt!]
 \centering
  \caption{Statistics of the obtained parameters over 1000 trials. Mean ($\mu$), standard deviation ($\sigma$) over mean, minimum and maximum. }
    \begin{tabular}{llrrrr}
    \toprule 
    \multicolumn{2}{l}{Par.} & \multicolumn{1}{c}{$\mu$ } & \multicolumn{1}{c}{$\sigma/\mu\cdot 10^2$} & \multicolumn{1}{c}{$min$} & \multicolumn{1}{c}{ $max$}\\
    \midrule
    $\varphi$&($^{\circ}$) &  33.99 & 0.079 & 33.94 & 34.21  \\
    $h_s$    &($GPa$)      &  4.03 &  7.303 & 3.15 &  5.11 \\
    $n$      &($-$)        &  0.30 &  1.262 & 0.28 & 0.31  \\
    $e_{c0}$ &($-$)        &  0.87 &  0.536 & 0.86  & 0.89  \\
    $\alpha$ &($-$)        &  0.15 &  2.540 & 0.14  & 0.16 \\
    $\beta$  &($-$)        &  1.44 &  2.153 & 1.32  & 1.55  \\
    \bottomrule
    \end{tabular}%
  \label{tab:varianza-par}
\end{table}%

Finally, to conclude the statistical analysis of the obtained result, we now focus on the correlation between all the parameters. The Pearson correlation coefficients between the full set of parameters is shown in Table \ref{tab:coef_va}, rounded to the the third digit.

\begin{table}[hbt!]
 \centering 
  \caption{Pearson correlation coefficient between the various parameters.}
    \begin{tabular}{lrrrrrr}
    \toprule 
    \multicolumn{1}{l}{ } &\multicolumn{1}{l}{$\varphi_c$} & \multicolumn{1}{c}{$h_s$} & \multicolumn{1}{c}{$n$} & \multicolumn{1}{c}{$e_{c0}$} & \multicolumn{1}{c}{ $\alpha$ } & \multicolumn{1}{c}{ $\beta$  }\\
    \midrule
     $\varphi_c$  & 1  & -0.053 & 0.001& -0.380 & 0.405& 0.451 \\
     $h_s$        &    &  1 \hspace{0.45cm}     & -0.911& 0.085& 0.038   & -0.210\\
     $n$          &    &        &1 \hspace{0.45cm}       & -0.322 &  0.165 & -0.130 \\
     $e_{c0}$     &    &        &       & 1 \hspace{0.45cm}      & -0.984 &  0.034 \\
     $\alpha$     &    &        &       &        & 1 \hspace{0.45cm}      &  0.075 \\
     $\beta$      &    &        &       &        &        &  1 \hspace{0.45cm} \\
    \bottomrule
    \end{tabular}%
  \label{tab:coef_va}%
\end{table}%

\begin{figure*}[htb]
  \centering
  \includegraphics[width=0.95\textwidth]{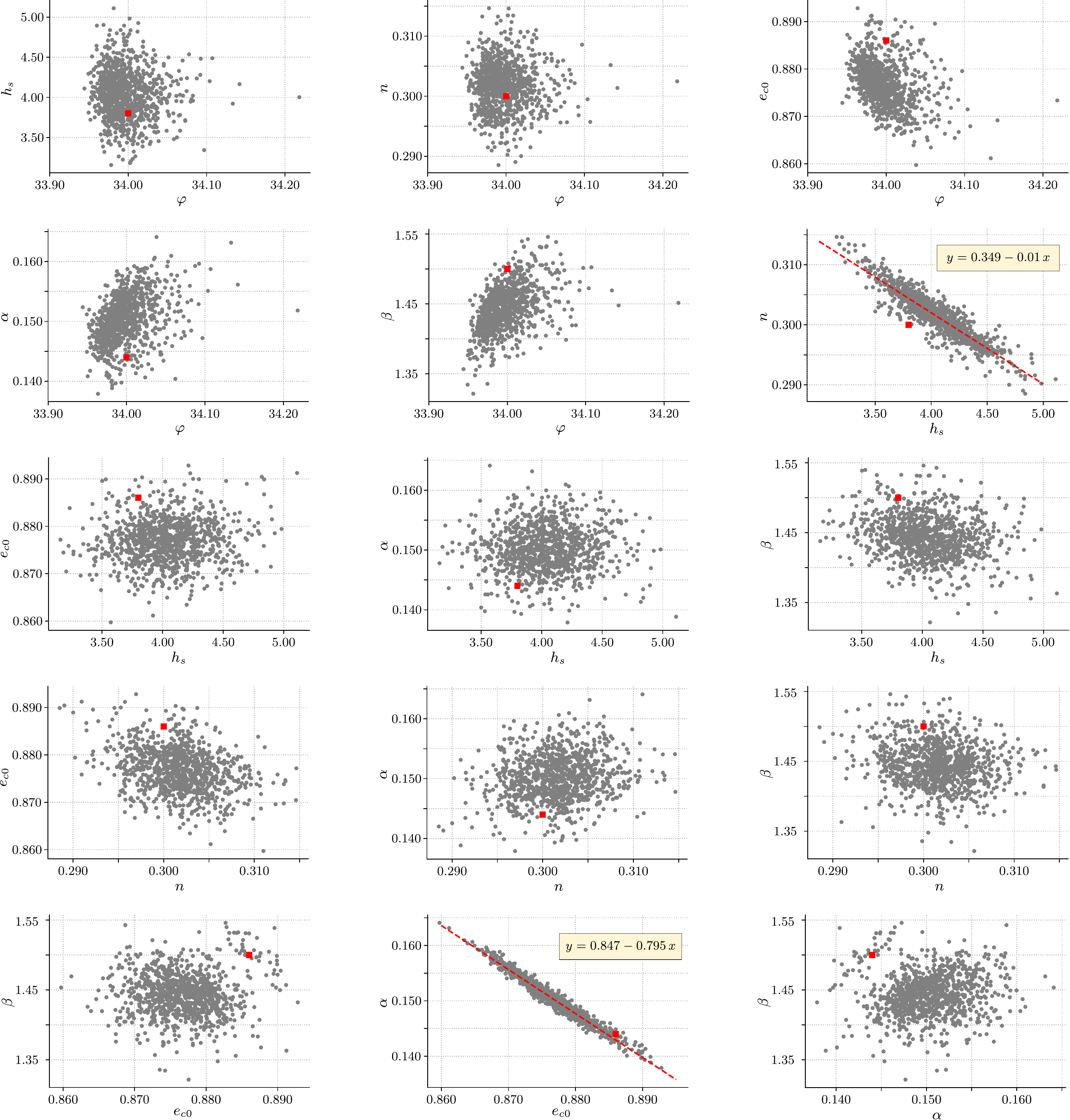}
  \caption{Collection of scatter plot showing the mutual distribution of different parameter pairs considering the results from the 1000 trials. The red square in each plot indicates the \emph{exact} solution. While most pairs are independent, the planes $(e_{c0},\alpha)$ and planes $(h_f-n)$ shows a linear trend. The equation from the linear regression is indicated in the corresponding plots.}
\label{fig:Hist_Convergenza1}
\end{figure*}

The correlation between the parameters $(h_s,n)$ and $(e_{c0},\alpha)$ is particularly evident. The full set of scatter plots describing the mutual distribution of parameters is shown in Figure \ref{fig:Hist_Convergenza1}. All the pairs of parameter that have low correlation are distributed with a polar symmetry around the peak. In each plot, the square marker indicates the position of the exact solution. For the correlated quantities, the equation for the linear regression is indicated in the corresponding plot. While the generalization of such linear trend outside the range of investigated properties requires additional investigations, it is important to observe that such a correlation reduces of the number of model parameters six to four. 

\subsection{Hochstetten sand calibration}\label{subpar:Calibrazione}

In the last simulation, the SH model is calibrated for the Hochstetten sand. The experimental data were obtained from the two oedometric tests and the three triaxial drained tests reported by von Wolfferdorff \cite{Wolffersdorff_A_hypoplastic_for_granular_material_with_a_predefined_limit_state_surface}. The initial conditions for this test are shown in the Table \ref{tab:cond_contorno}.

\begin{table}[hbt!]
 \centering
  \caption{Initial condition for the triaxial test (TxD1,TxD2,TxD3) e and the oedometer test (EDO1, EDO2) - von Wolefferdorff data \cite{Wolffersdorff_A_hypoplastic_for_granular_material_with_a_predefined_limit_state_surface}.}
    \begin{tabular}{lrrr}
    \toprule
    \multirow{2}{*}{Test} & \multicolumn{1}{c}{$T_1$ } & \multicolumn{1}{c}{$T_2$} & \multicolumn{1}{c}{$ e$ }  \\
       & \multicolumn{1}{c}{($kPa$)} & \multicolumn{1}{c}{($kPa$)} & \multicolumn{1}{c}{($-$)}  \\
    \midrule
    TxD1 & -100.0    & -100.0 &   0.690 \\
    TxD2 & -200.0    & -200.0 &   0.670 \\
    TxD3 & -300.0    & -300.0 &   0.660 \\
    EDO1 & -25.0     &  -12.5 &     0.730 \\
    EDO2 & -25.0     &  -12.5 &   0.695 \\
    \bottomrule
    \end{tabular}%
  \label{tab:cond_contorno}%
\end{table}%

The triaxial tests go on until the maximum deformation $\varepsilon_{fin}=0.20$ is reached, while the eodomeric tests continue until a void ratio $e_{fin}=0.672$ fore the EDO1 and $e_{fin}=0.643$ for the EDO2. 

We consider ratios $\lambda_d=0.60$ and $\lambda_i=1.20$  and the search interval indicated in table \ref{tab:Int_2}.

\begin{table}[hbt!]
 \centering
  \caption{Search space bounded vectors  $\mathbf{P}^*_{max}$and $\mathbf{P}^*_{min} $}
    \begin{tabular}{lrrrrrr}
    \toprule 
    \multirow{2}{*}{ID}&\multicolumn{1}{l}{$\varphi$ } & \multicolumn{1}{c}{$h_s$} & \multicolumn{1}{c}{$n$} & \multicolumn{1}{c}{$e_{c0}$} & \multicolumn{1}{c}{ $\alpha$ } & \multicolumn{1}{c}{ $\beta$  }\\
     &\multicolumn{1}{l}{($^{\circ})$} & \multicolumn{1}{c}{ (GPa)} & \multicolumn{1}{c}{(-)} & \multicolumn{1}{c}{(-)} & \multicolumn{1}{c}{ (-) } & \multicolumn{1}{c}{ (-) }\\
    \midrule
    max &40 & 9.0 & 0.40 & 1.1 & 0.20 & 2.0 \\
    min &25 & 1.0 & 0.25 & 0.6 & 0.05 & 0.9 \\
    \bottomrule
    \end{tabular}%
  \label{tab:Int_2}%
\end{table}%

The parameters of the GA are set to $ N_{i} = 500 $, $ N_I = 10 $, while the remaining parameters are taken as the default ones in algorithm \ref{alg:Genetic_Init.Pop} and \ref{alg:Genetic_Update.Pop}. 
The weights $ w_i $ $ i = 1,2,3 $ in the cost function \eqref{eq:Sco} are equal to unity, hence giving equal importance to the errors in each test.


The GA calibration provides the set of parameters shown third column (GA) of Table \ref{tab:W_H_GA_confronto}. This table also shows the values proposed by von Wolfferdorff (W)\cite{Wolffersdorff_A_hypoplastic_for_granular_material_with_a_predefined_limit_state_surface} and by Herel \& Gudehus (H) \cite{Herle_Gudehus_Determination_of_parameters_of_a_hypoplastic_constitutive_model_from_properties_of_grain_assemblies}.

\begin{figure}[h]
  \centering
  \subcaptionbox{Eodometric plane}[1\linewidth][c]{%
    \includegraphics[width=0.44\textwidth]{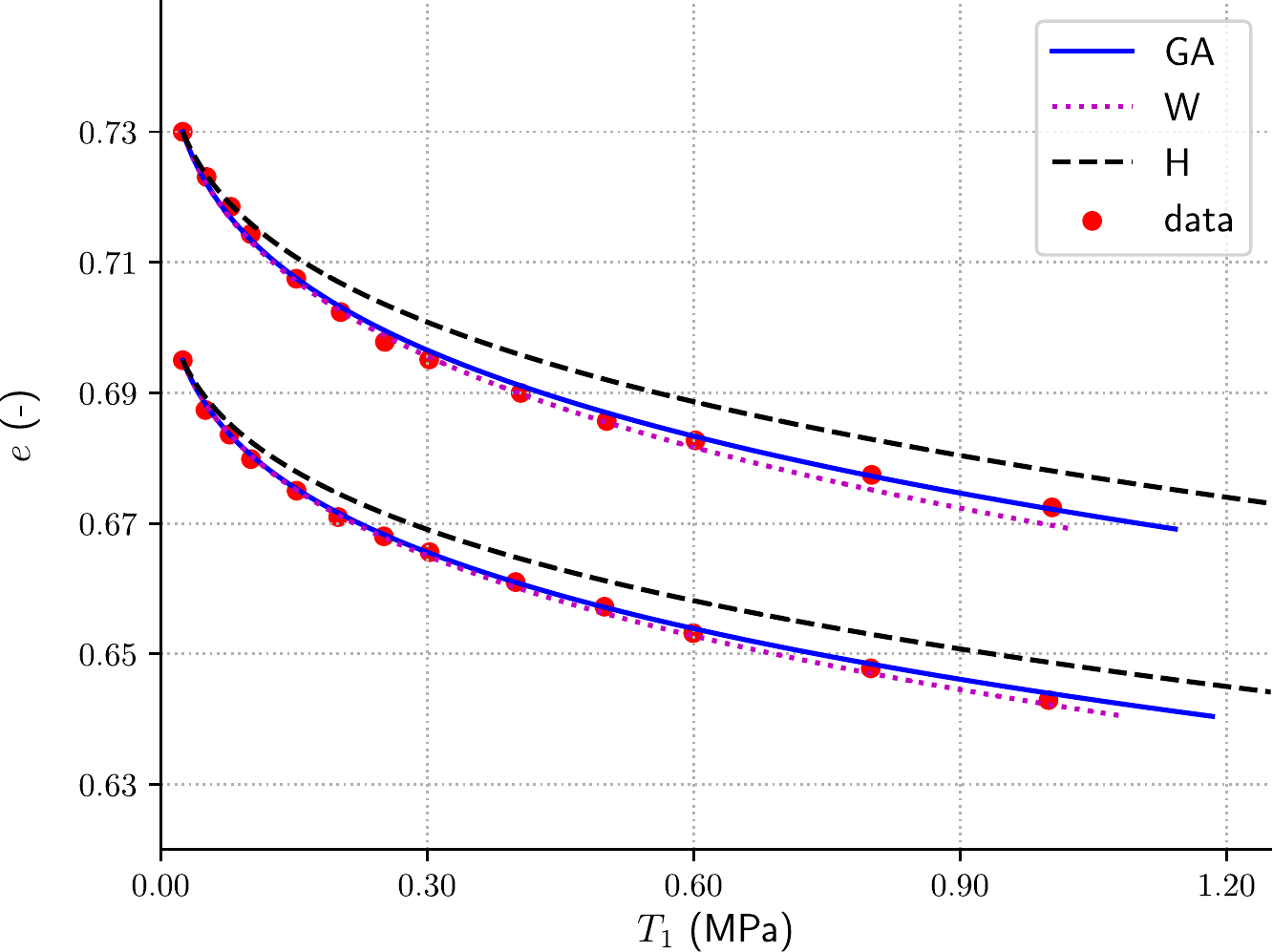}}\\
  \subcaptionbox{Triaxial deviatoric plane}[1\linewidth][c]{%
    \includegraphics[width=0.44\textwidth]{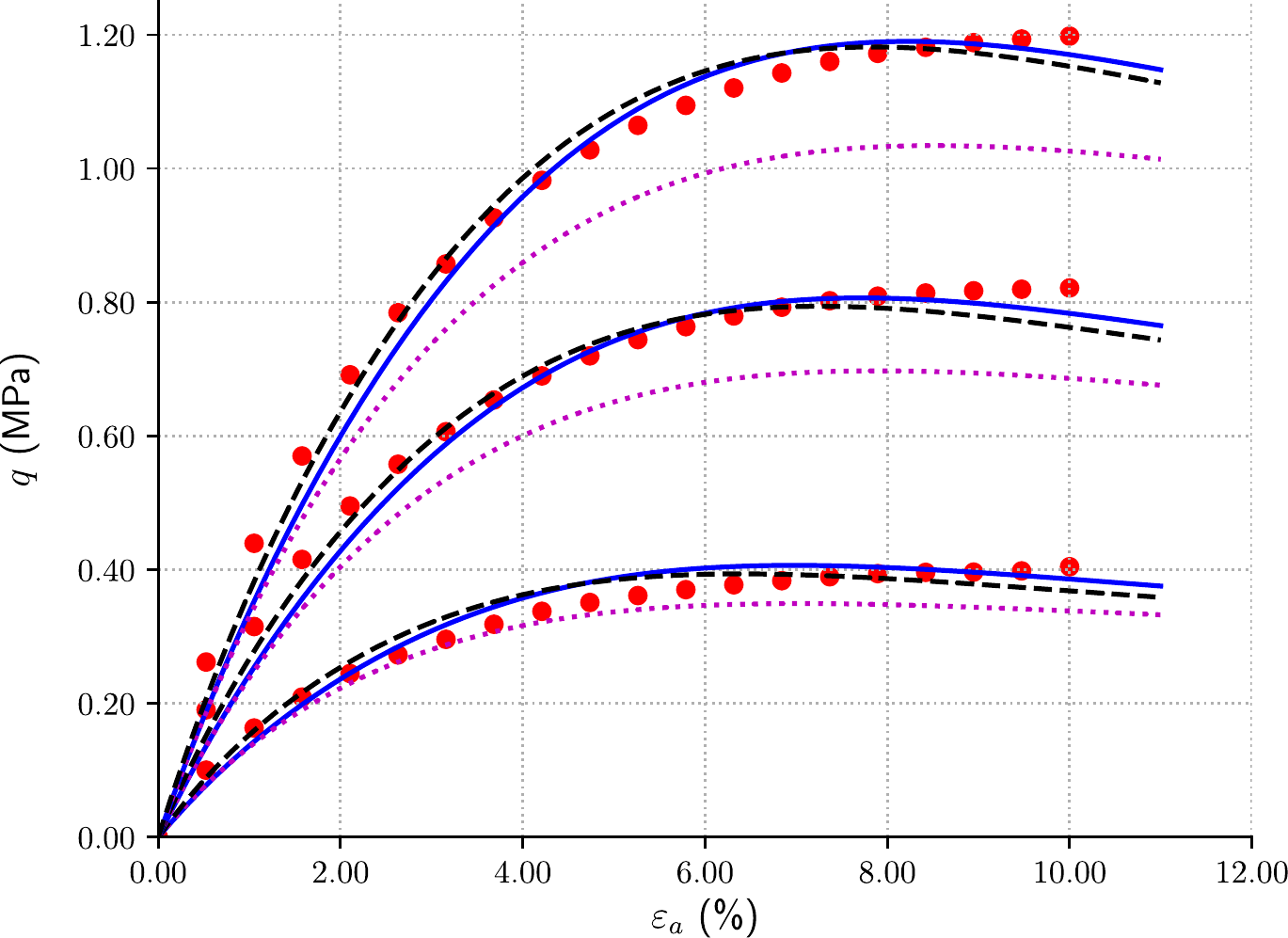}}\\
  \subcaptionbox{Triaxial volumetric plane}[1\linewidth][c]{%
    \includegraphics[width=0.44\textwidth]{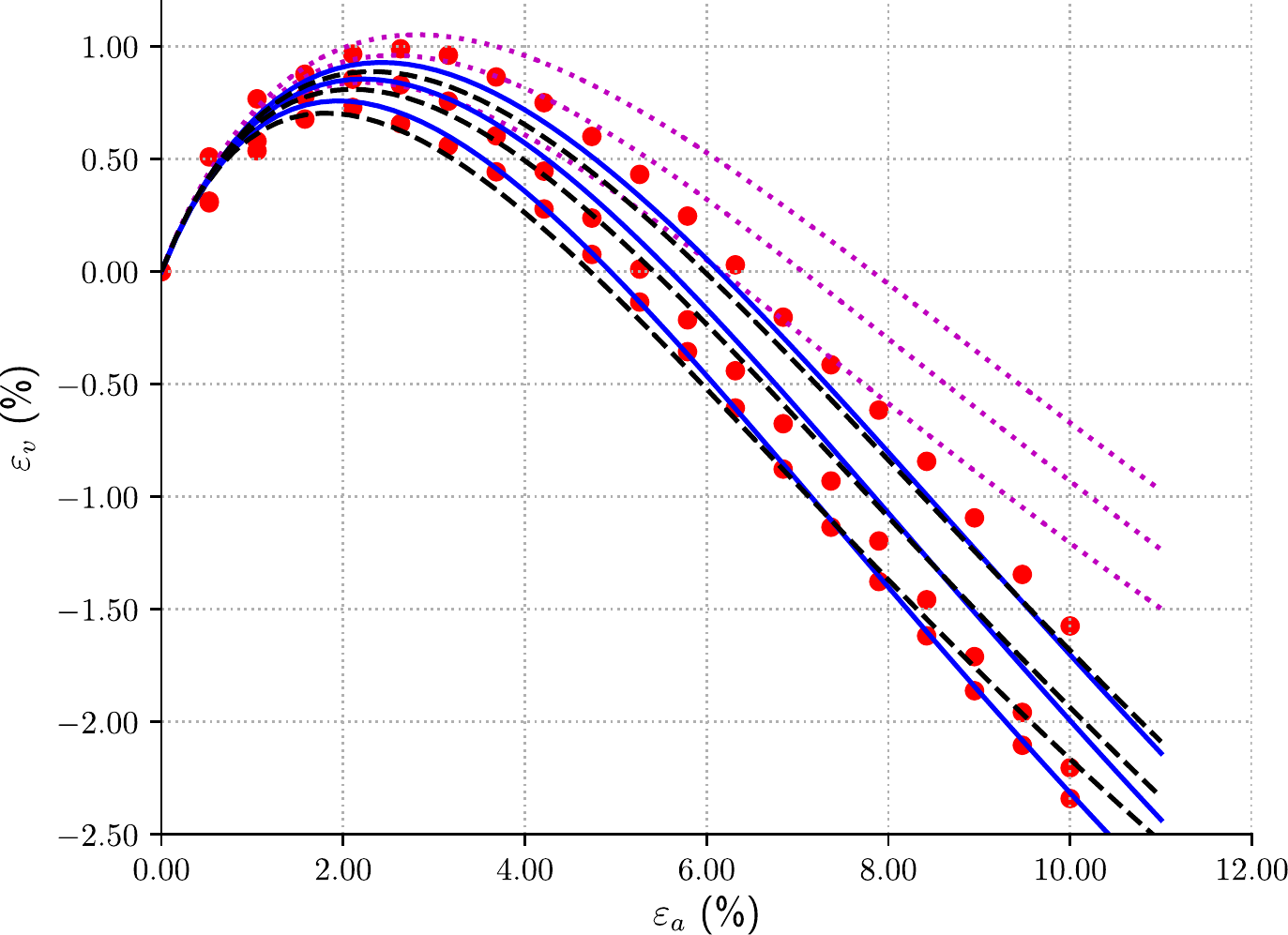}}
  \caption{Comparison between the response curves of the SH model and the experimental data (red point) on Hochstetten sand \cite{Wolffersdorff_A_hypoplastic_for_granular_material_with_a_predefined_limit_state_surface}. The magenta dotted line are computed using the parameters by von Wolfferdorff (W) \cite{Wolffersdorff_A_hypoplastic_for_granular_material_with_a_predefined_limit_state_surface}, the black hatched whit the parameters by Herel \& Gudehus (H) \cite{Herle_Gudehus_Determination_of_parameters_of_a_hypoplastic_constitutive_model_from_properties_of_grain_assemblies} and the blue solid whit the parameters obtained from the implemented procedure (GA). }
\label{fig:GA_Cal}
\end{figure}

The response curves calculated with these parameters are compared with the experimental data in the figure \ref{fig:GA_Cal}. The results show that the parameters suggested by von Wolfferdorff yields better description of the the oedometric response curves than what achievable using the parameters suggested by Herle \& Gudehus. The opposite is true in the regression  of the triaxial test, both in terms of volumetric deformations and deviatoric stresses.

The parameters obtained by the GA optimizer do not differ significantly from those proposed by the two authors. However, these yield better agreements in all the response curves, hence enabling better predictive capabilities of the SH model.

\begin{table}[h]
 \centering
  \caption{SH model parameter for the Hochstetten sand from von Wolfferdorff (W) \cite{Wolffersdorff_A_hypoplastic_for_granular_material_with_a_predefined_limit_state_surface}, Herel \& Gudehus (H) \cite{Herle_Gudehus_Determination_of_parameters_of_a_hypoplastic_constitutive_model_from_properties_of_grain_assemblies} and the GA calibration (GA).}
    \begin{tabular}{lcrrr}
    \toprule 
    \multicolumn{2}{l}{Par.} & \multicolumn{1}{c}{W} & \multicolumn{1}{c}{H} & \multicolumn{1}{c}{GA}\\
    \midrule
    $\varphi$&($^{\circ}$) & 33.00 &  33.00  & 32.73  \\
    $h_s$    &($GPa$)      &  1.00 &  1.50   & 1.32  \\
    $n$      &($-$)        &  0.25 &  0.28   & 0.23  \\
    $e_{i0}$ &($-$)        &  0.55 &  0.55   & 0.60  \\
    $e_{c0}$ &($-$)        &  0.95 &  0.95   & 1.04  \\
    $e_{d0}$ &($-$)        &  1.05 &  1.05   & 1.14  \\
    $\alpha$ &($-$)        &  0.25 &  0.25   & 0.23  \\
    $\beta$  &($-$)        &  1.50 &  1.00   & 1.26  \\
    \bottomrule
    \end{tabular}%
  \label{tab:W_H_GA_confronto}%
\end{table}%

\section{Conclusion}\label{Conc}

A novel procedure for the automatic calibration of the von Wolfferdorff's Sand Hypoplasticity (SH) model has been presented. The procedure is based on the solution of a regression problem in which the model parameters are adjusted so that a numerical model match experimental data. This data is provided by triaxial and eodometric tests, and the discrepancy between model prediction and experimental data is measured in dimensionless planes. The cost function is computed from the root mean square of the Fr\'{e}chet distances in these planes, and the regression is solved via Genetic Algorithms (GA).

After briefly reviewing the fundamentals of the SH model and their simplified formulation for the considered tests, the GA optimization is presented in detail. The numerical implementation of the SH model has been successfully validated using the popular ExCaliber-Laboratory Test Simulation.

A synthetic set of experimental datasets has then been used to study the relative importance, the uniqueness and the uncertainty of the parameters obtained by the GA calibration, and to explore their mutual correlation. Taking as benchmark test case hypothetical sand, the calibration has been repeated 1000 times, obtaining a large population of \emph{valid} sets of parameters. A statistical analysis of this population revealed that while the standard deviation of most of these is in the range 2 \%, the deviation in the granular hardness $h_s$ reaches up to 7\% of the expected value. These results highlight a minor impact of this parameter on the model. Furthermore, correlation analysis revealed that this parameter is linearly correlated with the parameter $n$. A strong linear correlation is also found for the parameters $e_{c0}\,-\,\alpha$. These results thus show that the set of parameters in the model can potentially be reduced.

Finally, the GA calibration is compared to the classical results from von Wolffersdorff, \cite{Wolffersdorff_A_hypoplastic_for_granular_material_with_a_predefined_limit_state_surface} and Herle \& Gudehus \cite{Herle_Gudehus_Determination_of_parameters_of_a_hypoplastic_constitutive_model_from_properties_of_grain_assemblies} on the Hochstetten sand. Overall, the proposed calibration yields better accuracy in matching the experimental data, enabling the automatic calibration within a few minutes of computation.

To conclude, the Genetic Algorithm calibration proved capable of correctly identifying the set of SH parameters from the experimental results of triaxial drained and eodometric compression tests. Moreover, the calibration allowed us to study the parameter uncertainty and their mutual correlation, paving the way towards data-driven reduction of the model parameters.

\FloatBarrier

\subsection*{Acknowledgements}
The authors gratefully acknowledge the support and the discussions with the engineer Pierantonio Cascioli, from GEINA srl, and Gabriele Sandro Toro, laboratory technician of the Department of Engineering and Geology of the Faculty Gabriele D'Annunzio of Chieti.

{\small
\bibliographystyle{ieee_fullname}
\bibliography{GA_SH_Calibration_Mendez_F_2020}
}

\end{document}